\documentclass[showpacs,amsmath,amstex,amssymb,mathfonts,pre]{revtex4-1}
\usepackage{graphicx}

\usepackage{tipa}
\usepackage{bm}

\usepackage{amsmath}
\usepackage{amssymb}
\usepackage{amsthm}
\usepackage{amsfonts}
\usepackage{float}
\usepackage{enumitem}
\usepackage{makecell}
\usepackage{multirow}
\usepackage{verbatim}

\begin{document}

\title{Efficient Intersection Control for Minimally Guided Vehicles: A Self-Organised and Decentralized Approach }

\author{Bo Yang and Christopher Monterola}
\affiliation{$^1$ Complex Systems Group, Institute of High Performance Computing, A*STAR, Singapore, 138632.}
\date{\today}

\date{\today}
\begin{abstract}
An important question for the practical applicability of the highly efficient traffic intersection control is about the minimal level of intelligence the vehicles need to have so as to move beyond the traffic light control. We propose an efficient intersection traffic control scheme without the traffic lights, that only requires a majority of vehicles on the road to be equipped with a simple driver assistance system. The algorithm of our scheme is completely decentralized, and takes into full account the non-linear interaction between the vehicles at high density. For vehicles approaching the intersection in different directions, our algorithm imposes simple interactions between vehicles around the intersection, by defining specific conditions on the real-time basis, for which the involved vehicles are required to briefly adjust their dynamics. This leads to a self-organised traffic flow that is safe, robust, and efficient. We also take into account of the driver comfort level and study its effect on the control efficiency. The scheme has low technological barrier, minimal impact on the conventional driving behaviour, and can coexist with the traffic light control. It also has the advantages of being easily scalable, and fully compatible with both the conventional road systems as well as the futuristic scenario in which driverless vehicles dominate the road. The mathematical formulation of our scheme permits large scale realistic numerical simulations of busy intersections, allowing a more complete evaluation of the control performance, instead of just the collision avoidance at the intersection.
\end{abstract}

\maketitle 
\section{Introduction}\label{sec_introduction}
The importance of traffic control at road intersections has been well recognized especially in large cities where the population and vehicle densities are high. The predominant methods of such control in modern cities utilise the traffic lights, and many schemes in optimizing such control are proposed. A single traffic intersection can be more efficiently controlled with either pre-signal systems\cite{cassidy} or with readily available sensing technologies, so that the traffic light can adapt to the real time demand of the intersection\cite{helbing,shamo}. Multi-layer agent control of a system of intersections in either a centralized or decentralized way has been intensively studied\cite{lammer,roozemond, bazzan, choy, cheu,silva,sun}. Effective vehicle to vehicle communications can also lead to traffic signals being optimized for connected vehicles\cite{ranka, meier}. More innovative approaches including cooperative driving where intersection manager sends out individual traffic signals to each vehicle are also reported\cite{wang, ahmane,wu,ahmane1}, so that the traffic flow across the intersection can be optimized with the sequence-based protocols. Retaining the  traffic light for the intersection control has the advantage of high robustness in terms of safety and ease of adaptation.

Research on signalized intersection control will continue to be of great importance in the foreseeable future in most parts of the world. On the other hand, in developed countries and mega-cities where the infrastructure is mature but strained, it is entirely possible that within 5$\sim$10 years a significant number of vehicles on the road will be autonomous or semi-autonomous\cite{footnote}.  In some of the technologically advanced cities the driverless cars can even become the mainstream method of transportation. Thus more recently, many researches have been focusing on lightless (with no traffic light) intersection control with autonomous vehicles in which the drivers are just passengers\cite{merge_algo,reservation_stone,rakha}. In these proposals, all the incoming vehicles relay information about their positions, velocities as well as their intentions to a central management system; such a central system sends out instructions to each of these vehicles, based on those information and an optimized algorithm\cite{luke,bazzan_opt}. In general such intersection control is vastly superior as compared to the signalized intersection control; on the other hand, a centralized intersection control also tends to be expensive for the infrastructure, especially when the number of vehicles it has to manage is large for busy intersections. More importantly, it requires near 100\% market penetration of autonomous vehicles, vehicles equipped with the adaptive cruise control, vehicle-to-vehicle as well as vehicle-to-infrastructure communications. Mixture of human drivers will again involve traffic lights as an auxiliary system\cite{reservation_stone}. 

It is, however, unlikely that the urban traffic is going to be dominated by the autonomous vehicles in the near future. We would anticipate a long transition period, during which the conventional vehicles coexist with autonomous or semi-autonomous vehicles. For transportation theory and engineering, this is thus an interesting period of time when we are on the verge of a major paradigm shift, while at the same time the inertia of conventional methods and technologies is equally great. Identification and development of a suitable set of tools and technologies that are compatible with both the new paradigm with the next generation technologies, as well as with the transition period before that, are thus very important, especially due to the uneven technological and economic progress across the world.

Faced with these challenges, in this paper we propose a new approach for the lightless intersection control that focuses on practicability, compatibility and upgradability, as well as ease of implementation with mature technologies. The National Highway Traffic Safety Administration (NHTSA) of the United States proposed a formal five-level classification of the autonomous vehicles\cite{nhtsa}. Our scheme only requires autonomous vehicles at Level 1 or above. Specifically, the vehicle only needs to be able to throttle or brake automatically based on the algorithms we have designed, in addition to the ability to communicate with the infrastructure at the intersection. Such ``feet-off" smart vehicles can be easily achieved with the mature technology, so we can focus on the lightless intersection control itself.

In particular, our scheme aims to cater to the following characteristics:

\begin{enumerate}
\item The proposed intersection control should offer significantly great advantages over various types of signalised intersection control, especially in terms of the intersection capacity. While such advantages will be gradual when the penetration of the ``feet-off" smart vehicles are low, there should be disruptive improvements in terms of the efficiency when such smart vehicles dominate the road.
\item The technological barrier of the new intersection control should be low.
\item The new intersection control should be able to accommodate a mixed traffic consisting of conventional vehicles with human drivers and the (semi-)autonomous ``feet-off" smart vehicles.
\item The new intersection control is completely distributed and decentralized; the agent at the intersection only collect and distribute information about the speed and position, while each vehicle decides its motion individually. Decentralized or distributed control is essential for reducing the infrastructure investment, vulnerability against attacks and system failures, as well as for allowing a mixed traffic.
\item The new intersection control can coexist with the conventional signalized control and can be adapted in stages.
\item The impact to the conventional driving experience is minimal.
\item The scheme architecture can be easily upgraded with more sophisticated algorithms that are commensurate with the sophistication of the hardwares.
\end{enumerate} 

Several schemes of the fully decentralized intersection control have been proposed in the literature\cite{negotiation, collision_avoidance, mini}. The focus has been on avoiding collisions in the intersection region via communication and negotiation between vehicles\cite{collision_avoidance}. Generally, vehicles negotiate for the right of way, especially for a few specifically defined spaces within the conflict zone: vehicles with the right of way can pass through the intersection, and collisions are avoided by allowing only one vehicle to be present in those pre-defined spaces at any instant. 

What is new in our scheme is that the impact of the intersection is completely modelled by the simple interactions between vehicles, just like the interactions between neighbouring vehicles travelling in the same lane on the highway. This naturally leads to a self-organised traffic flow through the intersection without the need of the traffic signals. As will be detailed in Sec.~\ref{algo}, we do not explicitly focus on the notion of the right of way in our decentralized scheme. The velocity and acceleration of the vehicles close to the intersection are updated real-time based on the vehicle's environment. Maintaining safe distances between any two approaching vehicles are thus achieved dynamically; only relative positions and velocities between vehicles are involved. In this way the issues of the precision of positioning and punctuality can be more easily managed. Our scheme also has the advantage of being significantly simpler and rather intuitive; in particular, the algorithm does not need to know the driver's intention explicitly. Only the few vehicles that are closest to the intersection will be involved in the communication and interaction, thus the processing power needed for each vehicle, as well as the communications between vehicles and the intersection agent, are minimal, allowing for very fast and robust reactions over potentially noisy channels with limited bandwidth. Our numerical simulation is based on the microscopic models instead of cellular automata, which is also much more realistic for empirical verification and practical implementations.

In addition to the above characteristics, we also would like to emphasize that by using the appropriate microscopic traffic model, our algorithm takes into full account of the vehicle-vehicle interaction, either due to the natural human behaviours or due to the intersection cruise control (a simple driver assistance device we will elaborate further in Sec.~\ref{design_hardware}). For high density traffic flows around busy intersections, the vehicle-vehicle interaction plays a very important role in simulating the impact of the intersection, even for vehicles far away from the intersection. This is especially true when the presence of the intersection induces traffic congestions and even traffic jams, which is by no means unusual. The simulations from our algorithm are thus more comprehensive and realistic. In particular, we can show explicitly the robustness and efficiency of the intersection control at very high inflow of the traffic. 

The paper will be organised as follows: In Sec.~\ref{design_hardware} we will introduce the overall design of our scheme; in Sec.~\ref{methodology} we give a detailed technical presentation on the methodology of our intersection control algorithm, focusing on the interactions between vehicles under various different situations; in Sec.~\ref{numerics} we show numerical simulations of the algorithm, both for vehicles equipped with the driver assistance device and for mixed traffic where a significant portion of the traffic is made of vehicles completely controlled by humans; in Sec.~\ref{op_com} we discuss the issue of safety, comfort and efficiency of the lightless intersection control based on our algorithm and numerical simulations; in Sec.~\ref{op_limit} we discuss some practical issues relating to the actual implementations and current limitations; in Sec.~\ref{summary} we summarize our results and discuss the outlooks and the related future works.
 
\section{General Design and Hardware Requirements}\label{design_hardware}

We will first introduce the main ideas in a heuristic manner. In our scheme, once a vehicle approaches the intersection, it receives relevant information about its environment. Based on these information each vehicle will decide its own behavior in a completely decentralized way, with a simple driver assistance device that may step in and decide the magnitude of the acceleration or deceleration based on the information and a predetermined set of rules. 

To achieve this, the first major change is to replace the traffic light with an agent for information relay: we call it the beacon (or a set of beacons), located at the intersection in place of the traffic lights. The beacon is able to detect the positions and velocities of the vehicles approaching the intersection within a certain range. It can also send the collected information to the vehicles approaching the intersection. 

The amount of information that can be collected and transmitted depends on the design and capability of the beacon. In the simple algorithm we are going to present in Sec.~\ref{algo}, we only require the beacon to collect the positions and velocities of two vehicles (moving towards the intersection) every lane that are closest to the intersection, as well as the vehicle that has just left the intersection for each direction. It also needs to pass such information to the same set of vehicles. For example, if there are $n$ lanes per road section in Fig.(\ref{schematics}) where the traffic is uni-directional, then in total the beacon will need to track the motions of at most $\sim 6n$ vehicles. A road section is defined as the segment of the road upstream of the intersection, in which all vehicles travel in the same direction. For uni-directional traffic, each intersection has two road sections; while for two-way traffic each intersection has four road sections.

If the vehicle is far away from the intersection, it is irrelevant to our intersection control. The control zone, as defined by the maximum distance stretching from the intersection (see Fig.(\ref{schematics})), is also called the interaction zone. The quantitative details of the interaction zone and their associated algorithm will be discussed in Sec.~\ref{algo}.

Another major component for the intersection control is the installation of a simple driver assistance device, which we call the intersection cruise control (ICC) device, on a conventional vehicle. The device is dormant when the vehicle is downstream of the intersection or far away from the beacon, and will only be switched on automatically when the vehicle is within the interaction zone. The ICC device, together with the programmed algorithms (see Sec.~\ref{algo}), is the key to ensure safety and robustness of the intersection control. 

The function of the ICC is two-fold. The first function is information relay, by either receiving information sent by the beacon at the intersection, or by communicating with nearby vehicles about the relative positions and velocities. This function is switched on whenever the vehicle is within the interaction zone and cannot be manually switched off. The second function is to compute the desired dynamics of the vehicle based on the information it receives and a set of predetermined rules. The ICC will enforce such dynamics by taking control over the acceleration and braking of the vehicle, but only when necessary and for a very brief period of time. The driver, however, can choose to override the control of the ICC anytime, but in the absence of human interference the ICC will guide the vehicle to move across the intersection, after which it will automatically switch off.

It is clear that if one considers autonomous or driverless vehicles, the ICC device only performs a very small set of tasks as compared to those technologically more advanced models. As will be clear from Sec.~\ref{algo}, the computational power required for each vehicle to make a real-time decision based on the given information is quite minimal. For our scheme of intersection control to work, only very basic vehicle-to-vehicle and vehicle-to-infrastructure communications are required. The number of vehicles involved in communication is also rather small, so from the technological point of view they can be easily implemented\cite{lunge}.

Apart from the communications, the ICC device does not need to be capable of other sensing abilities. In principle the active driver is still the human, who will be monitoring the road conditions and steering the wheel, so as to decide whether or not to override the ICC in uncommon situations. The ICC devices only control the acceleration and deceleration of the vehicle, with no control over the steering. This also makes the task of the ICC device rather simple.

The beacon can collect information about the positions and velocities in two ways. It can either mimic a lidar or radar system, using either laser or radio frequencies to track the vehicles approaching the intersection, or it can communicate with the vehicles, each of which can determine its own position and velocity fairly easily. A combination of both can also improve the reliability of tracking, especially when the traffic at the intersection is busy. Note the beacon itself does not do any computation to decide the dynamics of the vehicles. The vehicle-to-vehicle communication itself should be sufficient for all purposes of our control scheme; the presence of the beacon acts both as a way to enhance communication, as well as a fail-safe and monitoring device. 

\section{Methodology of Control and Modelling}\label{methodology}

Designing the optimal intersection control scheme depends both on the innovation of the control algorithm and the proper evaluation of the scheme performance via modelling and numerical simulation. Given that congestions are likely to occur due to the intersection, especially in urban areas with high vehicle density, an accurate modelling of the human driving behaviours and the impact of the intersections on those behaviours are important for the design and optimization of the traffic system. In this section we describe the general methodology employed in the intersection control and the modelling of the traffic system with the intersection.

Both the conventional intersection control and most of the innovative proposals in the literature have either a central control of the traffic at the intersection, or a set of complex protocols for the path selections and negotiations between vehicles sharing the intersection\cite{shamo,cassidy,sun,helbing,roozemond,bazzan,choy,cheu,silva,ahmane,wu,ahmane1,negotiation,mini,collision_avoidance}. One should note, however, in the absence of the intersection the traffic flows in a rather self-organised way, with each vehicle interacting mostly with its nearest neighbours, leading to many universal features and complex spatiotemporal patterns\cite{treiberbook,kernerbook}. Even though the details of such interaction are complex and still quite controversial\cite{helbingcrit,kernercrit,kernerbook}, the general forms are quite simple and can be modelled as follows\cite{yangbo1,dogbe,as}
\begin{eqnarray}\label{human}
a_n=f\left(h_n,v_n, \Delta v_n\right)
\end{eqnarray}
Here only the nearest neighbour interaction is considered, thus the driver's behaviour is only influenced by the headway and velocity of the vehicle right in front of him/her. In Eq.(\ref{human}) $a_n, h_n, v_n$ are the acceleration, bumper-to-bumper headway as well as the velocity of the $n^{\text{th}}$ vehicle in the lane, while $n$ increases in the direction of the motion of the vehicles. $\Delta v_n=v_{n+1}-v_n$ is the approach velocity of the $n^{\text{th}}$ vehicle. 

Though not commonly discussed in the literature, we expect Eq.(\ref{human}) to play a very significant role when the intersection is present, and it should be included in the modelling for the intersection control once the proper form of $f\left(h_n,v_n, \Delta v_n\right)$ capturing the essential human driving behaviours is determined. 

One should note that Eq.(\ref{human}) does not include the effect of the intersection. In our intersection control scheme, we extend such traffic models by treating the intersection as a source of perturbation. Instead of a central agent for the intersection control and/or sophisticated negotiation rules, we impose additional simple interactions to a small number of vehicles close to the intersection, and such interactions create a self-organised traffic flow even with the presence of the intersection, in which each individual vehicle governs its own dynamics, much the same as the vehicles travelling along the expressway. 
\begin{table}[]
\centering
\begin{tabular}{|l|l|lll}
\cline{1-2}
$S$                       & The set of all vehicles in the traffic system.    &  &  &  \\ \cline{1-2}
$S'$                      & The subset of vehicles close to the intersection. &  &  &  \\ \cline{1-2}
\multirow{2}{*}{$f_b$} & The boolean function defining the interaction     &  &  &  \\
                        & between vehicles in $S'$.                           &  &  &  \\ \cline{1-2}
$d_{S'}$                 & The deceleration function for vehicles in $S'$ when $f_b=1$.            &  &  &  \\ \cline{1-2}
\multirow{2}{*}{$f$} & Modeling of the human driving behaviour in     &  &  &  \\
                        & normal traffic flow (without intersections).                          &  &  &  \\ \cline{1-2}
\end{tabular}
\caption{The general architecture of the intersection control is completely defined by the five components in the table.}
\label{architecture}
\end{table}

Formally, let $n\in\{S\}$ be the index of the vehicles, where $S$ is the set of all the vehicles in the traffic system. We define $S'$ to be a small subset of $S$ consisting of just a few vehicles close to the intersection, and a specific example of the selection of $S'$ will be explained in Sec.~\ref{algo}. Additional interactions will be defined for vehicles with their indices $n\in\{S'\}$. To model such interaction, we define a Boolean function $f_b\left(x_n,v_n,\{x_{\bar n}, v_{\bar n}\}\right)$, where $x_n, v_n$ are positions and velocities of vehicles with $n\in\{S'\}$, and $\{x_{\bar n}, v_{\bar n}\}$ be the set of positions and velocities for all $\bar n\in\{S'\}$ and $\bar n\ne n$. $f_b$ takes the value of either $0$ or $1$. We thus generalise Eq.(\ref{human}), and the dynamics of each vehicle is thus defined by
\small
\begin{eqnarray}\label{human_extend}
a_n=\left\{
\begin{array}{lr}
d_{S'}\left(x_n,v_n\right)\quad n\in\{S'\}\text{ and }f_b\left(x_n,v_n,\{x_{\bar n},v_{\bar n}\}\right)=1\\
\qquad\qquad\qquad\text{ and }d_{S'}\left(x_n,v_n\right)<f\left(h_n,v_n,\Delta v_n\right)\\
f\left(h_n,v_n,\Delta v_n\right)\qquad\qquad\text{otherwise}\\
\end{array}
\right.
\end{eqnarray}
\normalsize
In Eq.(\ref{human_extend}) the first case is when the interaction between vehicles close to the intersection causes the driving behaviour of the $n^{\text{th}}$ vehicle to deviate from its normal driving behavior determined by Eq.(\ref{human}). Thus $d_{S'}\left(x_n, v_n\right)$ is generally negative, and its magnitude depends on the vehicle's position and velocity. The general model of Eq.(\ref{human_extend}) is the major contribution of this paper (also see Table.~\ref{architecture}), on which both the architecture of our intersection control scheme and the numerical simulations are based. We will now proceed to first determine the human driving behavoir modelled by $f\left(h_n,v_n, \Delta v_n\right)$ from the empirical data, followed by constructing a simple example of $S'$,  $d_{S'}\left(x_n,v_n\right)$ and $f_b\left(x_n,v_n,\{x_{\bar n},v_{\bar n}\}\right)$ that gives a particular set of algorithms for the self-organized, decentralized intersection control.

\subsection{Modelling Human Driving Behaviour}\label{humanmodel}

Though human driving behaviour can be modelled either macroscopically or microscopically, for practical applications we only focus on the microscopic models in the form of Eq.(\ref{human}). Many such models have been proposed in the past six decades\cite{treiberbook,kernerbook}, and this is still an active research area. We would like to emphasize, however, that the intersection control algorithm introduced in Sec.~\ref{algo} can accommodate various different types of traffic models as evidenced from our extensive numerical simulations, as long as the models reasonably reflect the main aspects of human driving behaviours. Here, it is possible for us to choose a simple traffic model that captures many essential features of the empirical spatiotemporal patterns. Such models are sufficient for the illustration of the efficiency as well as the robustness of the intersection control, as modelled by Eq.(\ref{human_extend}).

\begin{figure}[h!]
\begin{center}
\includegraphics[height=7cm]{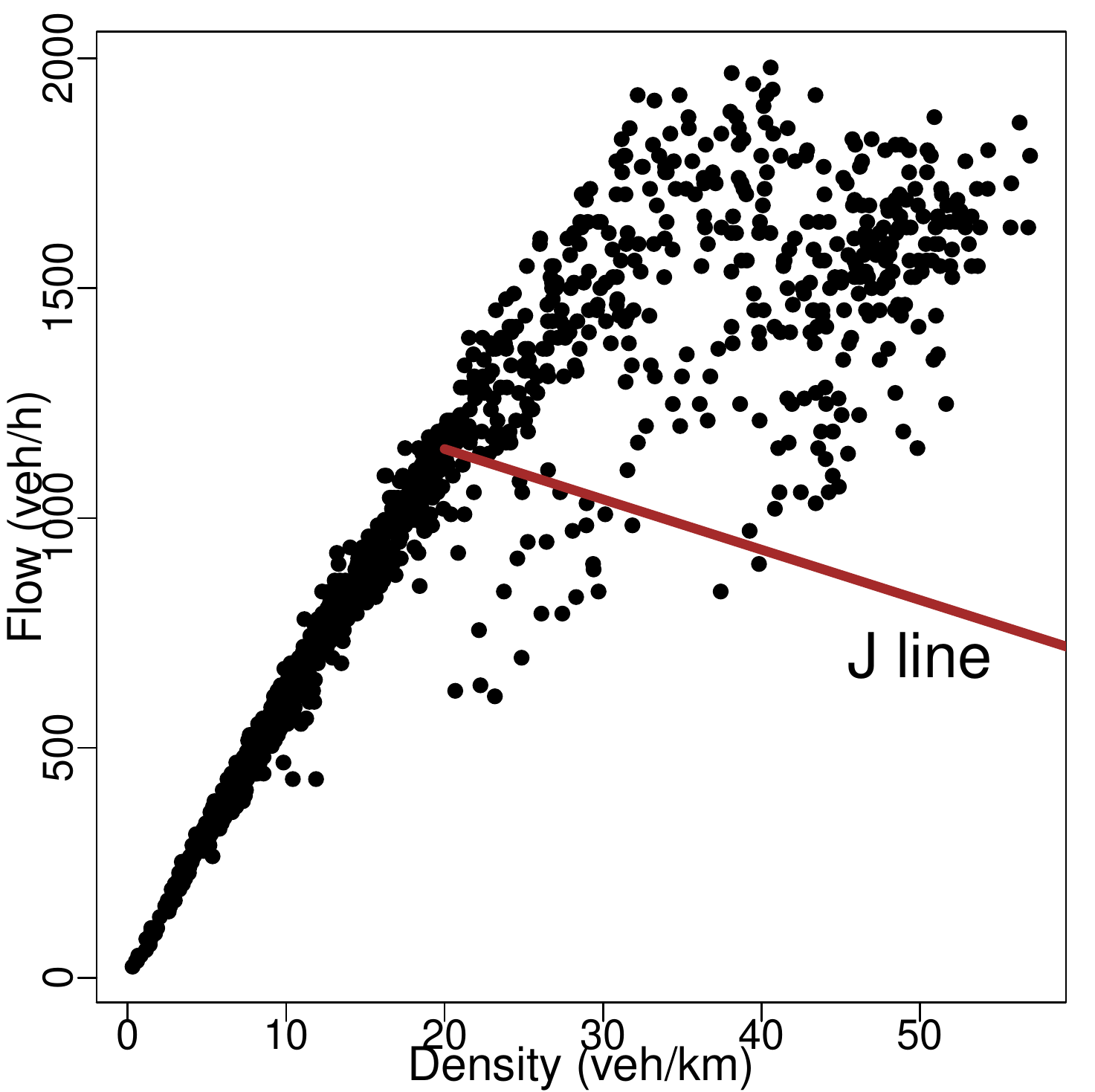}
\caption{The flow-density plot of the Queensway expressway traffic in Singapore, with empirical data taken from April 2015 to July 2015\cite{yangbo3}. Each data point is calculated from the traffic flow over a period of 5 minutes.}
\label{st}
\end{center}
\end{figure}

We will now follow the spirit of \cite{yangbo2} and outline the construction and tuning of the traffic model used in this paper, based on the universal features of the empirical flow-density plot. Such plot can be collected by measuring the flow and average velocities of the vehicles passing through a cross-section of the expressway/highway, and a typical flow-density plot from the Queensway expressway in Singapore is shown in Fig.(\ref{st}).

It can be seen clearly from Fig.(\ref{st}) that a phase transition occurs at density $\rho_{dj}\sim 20 veh/km$, and the metastable maximum flow rate $F_{\text{max}}\sim 2000 veh/h$ is reached at $\rho_c\sim 32 veh/km$. For $\rho<\rho_{dj}$ a fundamental diagram exists with a unique relationship between the flow and density. In the limit $\rho\rightarrow 0$, the gradient $\lim_{\rho\rightarrow 0}dF/d\rho$ gives the maximum velocity, or the speed limit of the traffic, $V_{\text{max}}\sim 80 km/h$. The wide moving jam that occasionally occurs during the peak hours give four characteristic quantities of the self-organized traffic flow, namely the velocity $V_j\sim 12 km/h$ of the jam moving upstream, the flow and density of the traffic downstream of the jam $F_{dj}\sim 1200 veh/h$ and $\rho_{dj}\sim 20 veh/km$, as well as the density of the traffic within the jam $\rho_j\sim 125 veh/km$, where the velocity of the traffic drops to zero. These four quantities give the $``J"$ line\cite{kernerbook} in Fig(\ref{st}). 

It is argued in \cite{yangbo2} that most (if not all) well-defined empirical features can be captured by the traffic model with the following form
\begin{eqnarray}\label{ovm}
a_n&=&f\left(h_n,v_n,\Delta v_n\right)\nonumber\\
&=&\kappa\left(V_{op}\left(h_n\right)-v_n\right)+\lambda_1\Delta v_n+\lambda_2|\Delta v_n|
\end{eqnarray}
where the optimal velocity function $V_{op}$ is tuned to fit a fundamental diagram with $V_{\text{max}}, V_{dj}, \rho_{dj}, F_{\text{max}}, \rho_c, \rho_{j}$. This is a modified asymmetric full velocity difference model\cite{afvd}, and specifically we have
\begin{eqnarray}\label{relationship}
&&V_{op}\left(\infty\right)=V_{max}, \quad V_{op}\left(h_{max}\right)=v_{dj}=F_{dj}/\rho_{dj}\nonumber\\
&&V_{op}\left(h_{min}\right)=0, \quad V_{op}\left(h_{cr}\right)=v_{cr}=F_{\text{max}}/\rho_{c}
\end{eqnarray}
The two other characteristic velocities from the flow-density plot are $V_j=F_{\text{max}}/\left(\rho_{c}-\rho_j\right)$, the velocity of the downstream front of a wide moving jam, and $V_C=\left(F_{\text{max}}-F_{dj}\right)/\left(\rho_{cr}-\rho_{dj}\right)$, the velocity of the downstream front between $F_{\text{max}}$ and $F_{dj}$. The characteristic headways are given by $h_{max}=\rho_{dj}^{-1}-l_c, h_{min}=\rho_j^{-1}-l_c, h_{cr}=\rho_{c}^{-1}-l_c$, where $l_c$ is the average length of the vehicle. The simplest way to construct such an optimal velocity function that satisfies Eq.(\ref{relationship}) is to make it continuous and piecewise. Defining $h_c=\frac{V_{max}-V_C}{v_{cr}-v_{dj}}\left(h_{cr}-h_{max}\right)-l_c$ we have:
\small
\begin{eqnarray}\label{piecewise}
V_{op}\left(h\right)=\left\{
\begin{array}{lr}
0 &  h<h_{min}\\
\frac{v_{cr}}{h_{cr}-h_{min}}\left(h+l_c\right)+V_j & h_{cr}>h\ge h_{min}\\
\frac{v_{cr}-v_{dj}}{h_{cr}-h_{max}}\left(h+l_c\right)+V_C& h_c>h\ge h_{cr}\\
V_{max} & h\ge h_c
\end{array}
\right.
\end{eqnarray}
\normalsize

The optimal velocity function is chosen to fit the important quantitative features of the fundamental diagram: it gives the correct maximum average velocity of the traffic system. It also gives the correct $(\rho_{dj}, F_{dj})$ and $(\rho_c, F_{max})$ pairs on the flow-density plot, where the traffic is still in the free flow phase. Note those two points are in general not collinear with the origin from the empirical measurements. In addition, it gives the maximum density $\rho_j$ of the traffic. While these are just some of the special points on the flow density plot, the tuning of the other parameters in the model ($\kappa_0, \lambda_1,\lambda_2)$ makes sure $(\rho_{dj},F_{dj})$ corresponds to the characteristic flow and density downstream of a wide moving jam, and $\rho_j$ corresponds to the density within a wide moving jam. One can also use the simplest form of $V_{op}$ that has been suggested in\cite{stepvop} for its analytic tractability; it however does not capture the correct fundamental diagram in the free flow phase. Another popular choice is the $V_{op}$ with a triangular fundamental diagram\cite{lighthill,treiberbook}, which assumes a single average velocity (close to the maximum velocity) at the free flow phase. While Eq. (\ref{piecewise}) is more realistic in the sense that it captures the characteristic velocity downstream of a wide moving jam, and a slowly decreasing average velocity in the free flow phase when the density increases towards the critical density, either form of the $V_{op}$ leads to similar numerical results for our intersection control simulations. 

The next step is to tune the parameters $\kappa, \lambda_1$ and $\lambda_2$ to capture the phase transition from the free flow phase to the congested phase. The latter is characterized by the scattering of data across a two-dimensional area in the flow-density plot. Due to the non-linear nature of the traffic model, the tuning can only be done numerically. In the unstable phase, cluster solutions will form\cite{yangbo2} from Eq.(\ref{ovm}) with two emergent extremal headways $h_{\text{max}}$ and $h_{\text{min}}$. With $h_{\text{max}}=\rho_{dj}^{-1}-l_c$ and $h_{\text{min}}=\rho_j^{-1}-l_c$, and the acceleration of the vehicles not exceeding $2ms^{-2}$, the three parameters can be fixed. Together with Eq.(\ref{piecewise}) the traffic model can be completely determined. The properly tuned parameters and the numerical simulation results will be shown in details in Sec.~\ref{numerics}.

One should note, however, the modified AFVD model here is known to be unrealistic when the consecutive vehicles have very different velocities, i.e. for a vehicle approaching far away towards a standing vehicle\cite{treiberbook}. For vehicles travelling across the intersection, such situation almost never happens. For simulations of accidents or immobilised vehicles at the intersection, the simple AFVD model needs to be generalised, or more realistic models (such as the intelligent driver model\cite{idm}) need to be used.

\subsection{Algorithms of the Intersection Cruise Control}\label{algo}

The paramount concern of the algorithm is the issue of safety. While accidents or disputes cannot be entirely eliminated, even with the traffic lights, the algorithm has to make sure that when such incidents do happen, the vehicle with the ICC device turned on will not be at fault. In this way, the active ICC device can also eliminate the chance of human errors and improve the overall safety when vehicles pass through the intersections.

\begin{figure}[h!]
\begin{center}
\includegraphics[height=5cm]{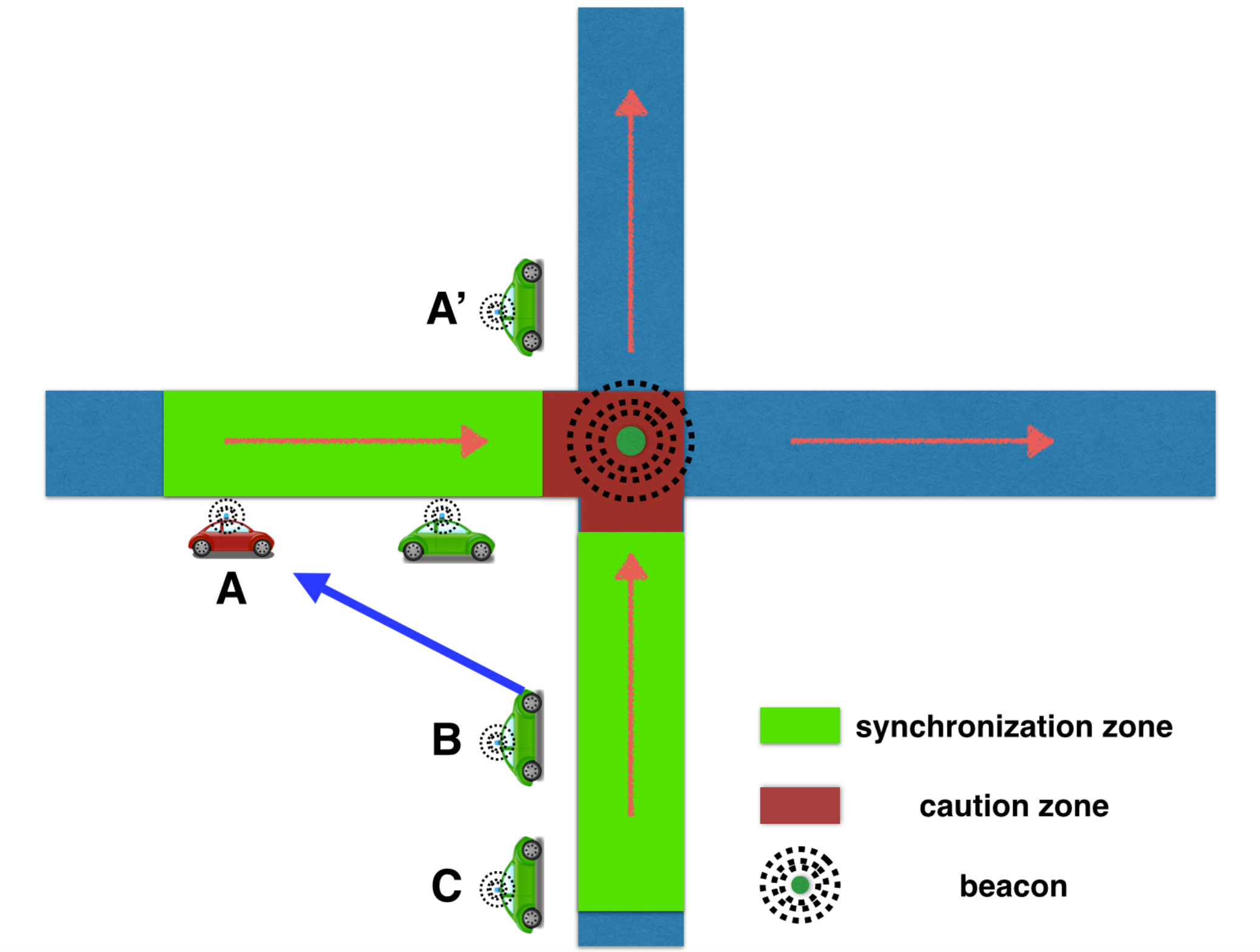}
\includegraphics[height=5cm]{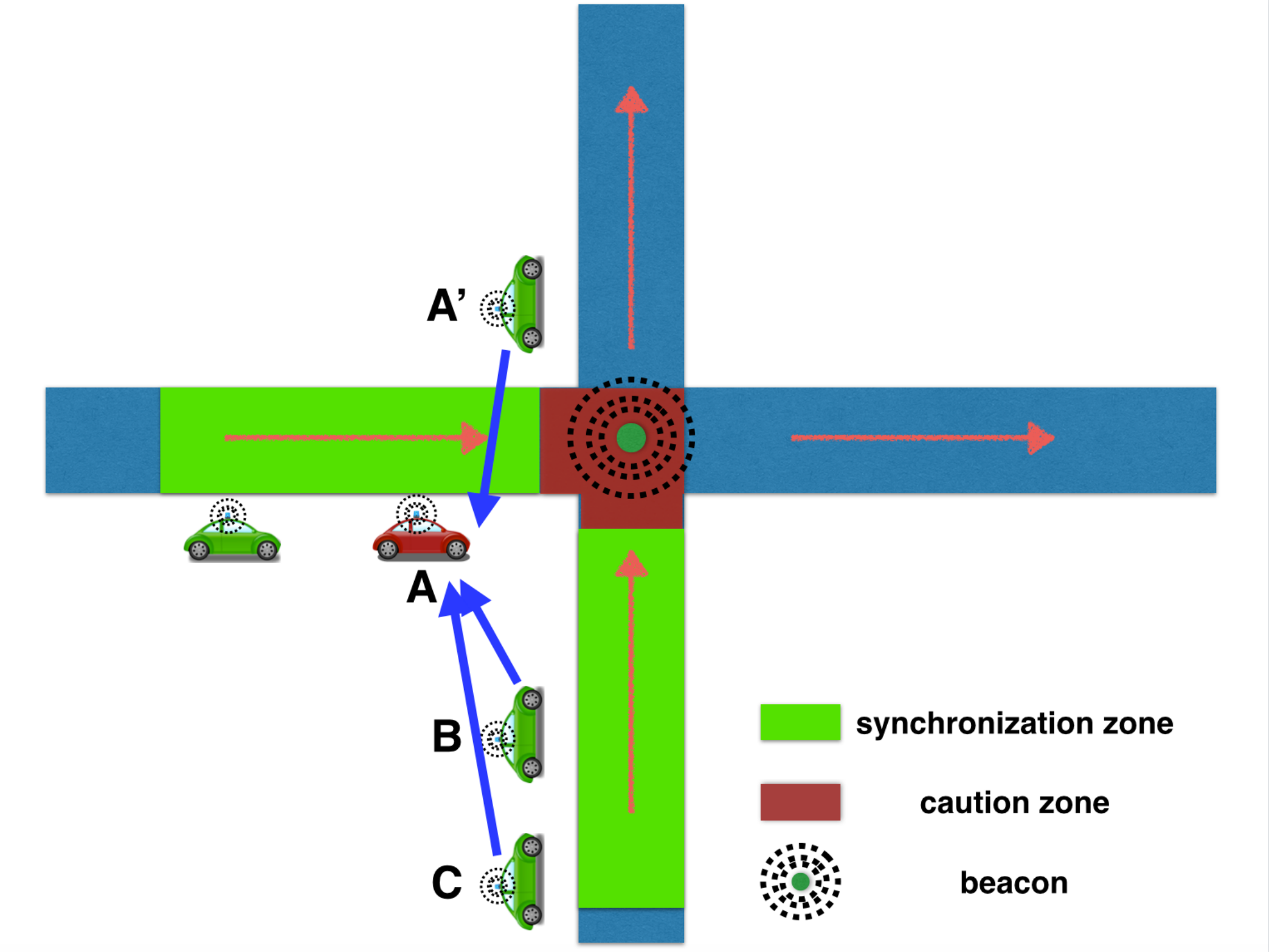}
\caption{Both in the left and right schematics, the red vehicle traveling from left to right is the Vehicle A described in Sec.~\ref{algo}. In the top schematics, Vehicle A only interacts with one vehicle, as directed by the blue arrow. In the bottom schematics, Vehicle A interacts with three vehicles as directed by the blue arrows.}
\label{schematics}
\end{center}
\end{figure}

We will now propose a specific set of algorithms in the form of Eq.(\ref{human_extend}), by properly defining $S', f_b$ and $d_{S'}$. One should note that for the other two components in Table.~\ref{architecture}, $S$ just contains all the vehicles in the simulation, while $f$ is already properly defined in Sec.~\ref{humanmodel}. We first propose a simple functional form of $d_{S'}$ by dividing the interaction zone into two parts: the caution zone, which starts at distance $L_c$ as measured from the intersection, and the synchronization zone, which starts at distance $L_s$ as measured from the end of the caution zone (see Fig.(\ref{schematics})). In addition to $L_s$ and $L_c$, the core parameters of the intersection control also includes $d_s$, the comfortable deceleration which is not exceeded when the vehicle is within the synchronization zone; $d_c$, the emergency deceleration not exceeded when the vehicle is within the caution zone; $V_m$, the speed limit of the road section. Now we propose the following simple example:
\begin{eqnarray}\label{deceleration}
d_{S'}\left(x,v\right)=\left\{
\begin{array}{lr}
d_c & L_c>x>0\\
d_s & \qquad L_c+L_s>x>L_c\\
\end{array}
\right.
\end{eqnarray}
where $x$, the position of the vehicle, is also measured from the intersection. Thus depending on where the vehicle is in the interaction zone, it takes up the constant deceleration of either $d_s$ or $d_c$ when its normal driving behavior is interrupted due to the interaction with vehicles travelling in the other direction towards the intersection. In this simple example, we make the deceleration independent of the vehicle's velocity.

The core parameters mentioned above, however, are not independent. The length of the caution zone is given by
\begin{eqnarray}\label{cautionzone}
L_c=\frac{V_m^2}{2d_c}
\end{eqnarray}
which basically implies that the vehicle entering the caution zone always has enough distance to decelerate comfortably to zero velocity before it reaches the intersection, as long as the vehicle does not exceed the speed limit $V_m$. Here, $V_m, d_s$ and $d_c$ are set by the configuration of the road system, local traffic laws as well as the driver's comfort level, and we treat them as fixed parameters. The length of the synchronization zone, $L_s$, on the other hand, is tunable.

For illustration purpose in this section we look at the simplest type of the traffic intersection, where the two road sections are uni-directional and each consisting of only one lane. We also do not allow turning of the traffic. The extension to the more complex road system is straightforward, and we will discuss it in Sec.~\ref{numerics} and Sec.~\ref{summary} . 

The set of vehicles $S'$ is illustrated in Fig.(\ref{schematics}). The interaction zone is upstream of the intersection for each lane; for each lane, only the two vehicles closest to the intersection within the interaction zone, together with the vehicle that just passed the intersection, are included in $S'$. Formally for each lane, let $x_n$ be the position of the $n^{\text{th}}$ vehicle as measured from the intersection. $n$ increases in the direction of the traffic flow, while $x_n$ increases upstream of the intersection. $n\in S'$ if and only if one of the following conditions are satisfied:
\begin{eqnarray}\label{sset}
&&0<x_n<L_s+L_c\text{ and }x_{n+2}<0\\
&&x_n<0\text{ and }x_{n-1}>0
\end{eqnarray}
Note however only vehicles upstream of the intersection, i.e. $x_n>0$, will be governed by Eq.(\ref{human_extend}).

With the proper definition of $S'$, we will proceed to define $f_b$. We denote the road section where vehicles travel from west to east as Section 1, while the other section with vehicles going from south to north as Section 2. Since the intersection control is completely decentralized, it is sufficient to look at the behaviour of any one vehicle. Without loss of generality we will focus on a single vehicle approaching the intersection in Section 1, traveling from west to east. We denote it as Vehicle $A$ (see Fig.(\ref{schematics})).

We first start with a heuristic description. In the absence of the intersection, the dynamics of this vehicle will be completely determined by Eq.(\ref{human}), which mimics the natural human driving behavior by following the preceding vehicle in an adaptive way. Our algorithm is very straightforward: the only effect of the presence of the intersection is that when certain specific criteria are satisfied for Vehicle A, the vehicle is obliged to decelerate (as specified in Eq.(\ref{human_extend}) and Eq.(\ref{deceleration})). The dynamics of the vehicle is thus completely defined by an unambiguous specification of those criteria. In road section 2, we denote the vehicle closest to the intersection but moving away from the intersection as Vehicle $A'$, the vehicle closest to the intersection but approaching the intersection as Vehicle $B$, and the vehicle right behind Vehicle $B$ as Vehicle $C$ (see Fig.(\ref{schematics})).

The criteria for Vehicle $A$ to decelerate are broadly divided into two cases. The first case is when there is exactly one vehicle in front of Vehicle $A$ that is also approaching the intersection (Top figure of Fig.(\ref{schematics})). In this case Vehicle $A$ will only interact with Vehicle $B$. Let their respective distances from the intersection be $l_A, l_B$, and their respective velocities be $v_A, v_B$. The two intersection time headways are thus given by $t_A=l_A/v_A, t_B=l_B/v_B$. We also define the safe intersection time headway $\bar t_B=l_{\text{safe}}/v_B$. Vehicle A will undergo deceleration when 
\begin{eqnarray}\label{i1}
\text{C1}:\qquad t_A>t_B\text{ and } t_A-t_B<\bar t_B+t_{\text{safe}}
\end{eqnarray}
Here we introduce two safety parameters $l_{\text{safe}}$ and $t_{\text{safe}}$. Physically, the front bumper of Vehicle $B$ has to be at least a distance of $l_{\text{safe}}$ away from the intersection, at the time when Vehicle $A$ enters the intersection. This gives enough margin to prevent the two vehicles from colliding. The value of $l_{\text{safe}}$ will thus be determined by the width of the intersection and the length of the vehicle. $t_{\text{safe}}$ is the safety allowance to make sure that $t_A-t_B<l_B/v_B$ holds strictly throughout the period when Vehicle $A$ approaches and passes the intersection. In the ideal case one can set $t_{\text{safe}}=0$, and in general one should set a small positive value for $t_{\text{safe}}$ to account for finite resolution/accuracy of either the actual implementation or numerical simulation. We will also show later that increasing $t_{\text{safe}}$ can make the control algorithm more robust.

The second case is when Vehicle A is the leading vehicle in road Section 1 that approaches the intersection (Bottom figure of Fig.(\ref{schematics})). In this case Vehicle $A$ will interact with Vehicle $A'$, Vehicle $B$ and Vehicle $C$, with distances to the intersection given by $l_{A'}, l_B$ and $l_C$, and the velocities given by $v_{A'}, v_B$ and $v_C$ respectively. Note here we take $l_{A'}$ to be negative since Vehicle $A'$ has already passed the intersection. The intersection time headways for Vehicle $B$ and Vehicle $C$ are given by $t_B=l_B/v_B$ and $t_C=l_C/v_C$. The safe intersection time headways are $\bar t_{A'}=\left(l_{A'}+l_{\text{safe}}\right)/v_{A'}, \bar t_B=l_{\text{safe}}/v_B$ and $\bar t_C=l_{\text{safe}}/v_C$. The conditions for Vehicle $A$ to undergo deceleration are any one of the following
\begin{eqnarray}
&&\text{C2}:\qquad t_A>t_B\text{ and } t_A-t_B<\bar t_B+t_{\text{safe}}\label{i2}\\
&&\text{C3}\qquad t_A>t_C\text{ and } t_A-t_C<\bar t_C+t_{\text{safe}}\label{i3}\\
&&\text{C4}\qquad \bar t_{A'}>0\text{ and }\bar t_{A'}<t_A\label{i4}
\end{eqnarray}

It is easy to summarize all these conditions by the formal definition of $f_b\left(x_n,v_n,\{x_{\bar n},v_{\bar n}\}\right)$, since conditions C1$\sim$C4 are just functions of $x_n,v_n$ with $n\in S'$. We thus have
\begin{eqnarray}\label{fb}
&&f_b\left(x_n,v_n,\{x_{\bar n},v_{\bar n}\}\right)\nonumber\\
&&=\left\{
\begin{array}{lr}
1 & x_{n+1}>0\text{ and C1}\\
1 & \qquad x_{n+1}<0\text{ and }\left(\text{C2 or C3 or C4}\right)\\
0 & \text{otherwise}
\end{array}
\right.
\end{eqnarray}

In the synchronization zone, C1$\sim$C3 make sure that the intersection time headway difference between Vehicle $A$ and either Vehicle $B$ or Vehicle $C$ is always slightly larger than the safe intersection time headway of the vehicle that arrives at the intersection first (thus we use the term ``synchronization"). C4 makes sure that Vehicle A will not collide with the vehicles in the other road section, whose head bumper has already passed the intersection. 

When Vehicle $A$ is in the caution zone and one of the conditions of C1$\sim$C4 is satisfied, the vehicle will brake with deceleration $d_c$. This can happen due to the decentralized nature of the intersection control and the length of the synchronization zone may not be long enough, especially when the density of the traffic is high. Additional algorithms in the caution zone could be implemented to further enhance the robustness of the algorithm, and this will be explored in future works. For the simple intersection we are looking here and some of the more sophisticated intersections in Sec.~\ref{numerics_phase}, the current scheme is robust enough.

\begin{table*}[!ht]
\begin{tabular}{|l|l|l|l|}
\hline
Parameters & Numerical value          & Physical meaning                                                                                                    & Comments                                                                                                                                                                                                  \\ \hline
$V_m$       & $80 km/h$                  & \begin{tabular}[c]{@{}l@{}}Maximum velocity of the \\ traffic/speed limit.\end{tabular}                             & \multirow{4}{*}{\begin{tabular}[c]{@{}l@{}}These are the core parameters of the intersection control, \\ that are fixed by road design, driver comfort level \\ as well as local traffic rules.\end{tabular}} \\ \cline{1-3}
$L_c$       &           $25m$               & \begin{tabular}[c]{@{}l@{}}Length of the caution zone.\end{tabular}         &                                                                                                                                                                                                           \\ \cline{1-3}
$d_s$       & $2 ms^{-2}$  & \begin{tabular}[c]{@{}l@{}}Maximum deceleration in the synchronization \\ zone.\end{tabular}                        &                                                                                                                                                                                                           \\ \cline{1-3}
$d_c$       & $5 ms^{-2}$ & Maximum deceleration in the caution zone.                                                                           &                                                                                                                                                                                                           \\ \hline
$L_s$       &      $40\sim 60m$                    & \begin{tabular}[c]{@{}l@{}}Length of the synchronization zone.\end{tabular} & \begin{tabular}[c]{@{}l@{}}This is a tunable core parameter of the intersection \\ control\end{tabular}                                                                                                   \\ \hline
$l_{\text{safe}}$    & $10$ m                     & \begin{tabular}[c]{@{}l@{}}Safe distance between vehicles passing \\ through the intersection.\end{tabular}         & \multirow{2}{*}{\begin{tabular}[c]{@{}l@{}}These two are safety parameters to \\ prevent collisions at the intersection.\end{tabular}}                                                                    \\ \cline{1-3}
$t_{\text{safe}}$    & $0.1$ s                    & \begin{tabular}[c]{@{}l@{}}Time allowance to improve the robustness \\ of the algorithm.\end{tabular}                   &                                                                                                                                                                                                           \\ \hline	
\end{tabular}\label{t2}
\caption{Various parameters used in the algorithm and their physical interpretations}
\end{table*}
One should note that when $f_b=0$, the dynamics of Vehicle $A$ is simply governed by Eq.(\ref{human}) and in most cases the vehicle accelerates. Thus when the intersection time headway of Vehicle $A$ is smaller than that of either Vehicle $B$ or Vehicle $C$, it tends to accelerate so as to make sure when it passes the intersection the vehicles from the other road section are a safe distance away. Vehicle $B$ or Vehicle $C$ are designed to decelerate if necessary to guarantee safety.

If the intersection time headway of Vehicle $A$ is larger than that of Vehicle $B$ or Vehicle $C$, and if the difference of the time headway is larger than $\bar t_B$ or $\bar t_C$, Vehicle A will also tend to accelerate if the vehicle in the front is not too close (Eq.(\ref{human}) guarantees that), since it is safe to do so. The algorithm tends to minimize unnecessary deceleration of Vehicle $A$ to improve the efficiency. In the synchronization zone, the vehicles travelling in different directions approaching the intersection will tend to synchronize their intersection time headway to be just nice for them to pass each other safely, leading to a self-organized flow across the intersection. 

It is possible, though extremely unlikely, for Vehicle A and Vehicle B in the right schematics of Fig.(\ref{schematics}) to be of exactly the same distance away from the intersection, travelling at exactly the same velocity. In this situation of exact symmetry, the algorithm above will decelerate both vehicles equally until they come to a full stop, which is pathological. This, however, can be easily solved by randomly select Vehicle A or B and let it keep its velocity for one or a few time step. In this way the exact symmetry is broken. The algorithm makes sure there will be no exact symmetry in the subsequent time evolution, and the vehicles will pass through the intersection smoothly.

\section{Numerical Simulation}\label{numerics}

In this section, we perform the numerical simulation of the simplest road configuration to illustrate the feasibility of the algorithm introduced in the previous section. In most cases, a properly tuned optimal velocity (OV) model is sufficient enough to model the human drivers, which is tuned as described in Sec.~\ref{humanmodel}. The parameters in the model are $\kappa=0.1 s^{-1}, \lambda_1=0.39 s^{-1},\lambda_2=-0.2 s^{-1}$, and the optimal velocity function is given by
\begin{eqnarray}\label{ov}
V_{op}\left(h\right)=\left\{
\begin{array}{lr}
0 &  h<3m\\
\kappa_1\left(h-h_{min}\right)& 27m>h\ge 3m\\
\kappa_2h+v_0& 56m>h\ge 27m\\
V_{max} & h\ge 56m
\end{array}
\right.
\end{eqnarray}

Here $\kappa_1=0.71s^{-1}, \kappa_2=0.17s^{-1}, h_{min}=3m, v_0=12.4ms^{-1},V_{max}=22ms^{-1}$. Eq.(\ref{ov}) makes sure the fundamental diagram of the traffic flow in the free flow phase agrees with the empirical data quantitatively. The parameters in Eq.(\ref{ovm}) also captures the empirical phase transition from the free flow phase to the congested phase. To be consistent with Eq.(\ref{ov}), the maximum velocity is set by $V_m=22ms^{-1}$ . While in principle the speed limit $V_m$ can also be tuned freely, it strongly affects the human driving behaviours and thus $f\left(h,v,\Delta v\right)$. We will keep $V_m$ fixed here to focus on the algorithm of the intersection control.

The values of the fixed core parameters of the algorithm are listed in Table. II. Since in our simple case we do not allow turning of the vehicles at the the intersection, each section of the road just represents a single lane traffic that is externally perturbed by the intersection. Such a traffic flow with the external perturbation is well understood within the framework of car following models\cite{bando}. Depending on the density of the traffic, the free flow of the traffic could be stable, metastable or linearly unstable against perturbations. 

\subsection{Phases of the Intersection Flow}\label{numerics_phase}

With the intersection control scheme we proposed, the phases of the intersection flow is closely related to the phases of the highway traffic flow\cite{bando,kerner}. Here the traffic flow is perturbed by the intersection, and depending on the density of the traffic and the strength of the perturbation, the intersection may or may not induce traffic congestions.

\begin{figure}[h!]
\begin{center}
\includegraphics[height=8.5cm]{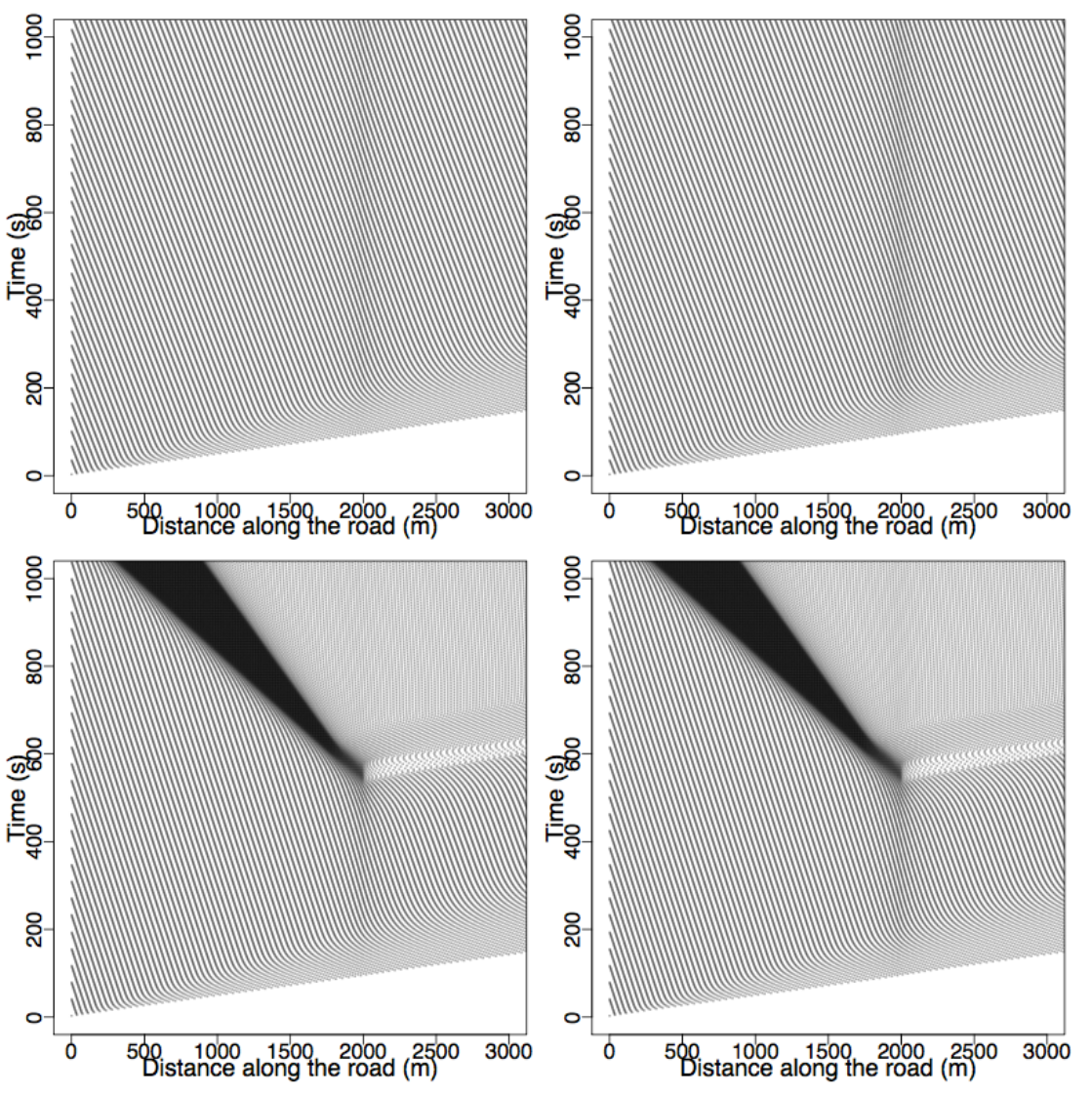}
\caption{The spatio-temporal representations of the traffic flow. The x-axis is the distance along the road, while the y-axis is the time of simulation. The intersection is located at $2000 m$ for both the road sections. Each black dot in the plot represents a vehicle, located at a specific time and position. The top two plots represents the free flow phase, where the plot on the left is the traffic flow of the road section 1, while the one on the right is the traffic flow of the road section 2. The bottom two plots show the congested flow in both road section 1 (on the left) and road section 2 (on the right). We can see clearly a dense region of wide moving jams with growing width moving upstream. The growing width is due to the fact that the inflow of the traffic at the upstream front of the jam is greater than the outflow of the traffic at the downstream front.}
\label{phase}
\end{center}
\end{figure}

In the free flow phase, the traffic in-flow $F_{\text{in}}$ equals to the out-flow $F_{\text{out}}$ of the intersection, thus by definition the intersection capacity $C$ is greater than $F_{\text{in}}$. In the congested phase, however, there is an accumulation of vehicles in the traffic system due to the presence of the intersection, thus $F_{\text{in}}>C$ and for traffic system with intersection we can define the two phases as follows:
\begin{eqnarray}
F_{\text{in}}&<&C\qquad\text{         Free flow phase }\nonumber\\
F_{\text{in}}&>&C\qquad\text{         Congested phase}\nonumber
\end{eqnarray} 
In all the numerical simulations carried out in this paper, the vehicles are inserted randomly into the traffic system with a constant average flow rate. The spatio-temporal distribution of the vehicles in these two phases are plotted in Fig.(\ref{phase}), where we assume all vehicles are equipped with the ICC device. Since multiple OD pairs exist for a single intersection, the congested phase is defined as the case when at least one OD pair has the inflow larger than the outflow. Note that by our definition, the hallmark of the congested phase is the emergence of the wide moving jams at the intersection that move upstream. Moderate congested traffic may be present at the intersection, but in the free flow phase congested traffic is always confined near the intersection and dissipates upstream of the intersection.

\begin{figure}[h!]
\begin{center}
\includegraphics[height=8.5cm]{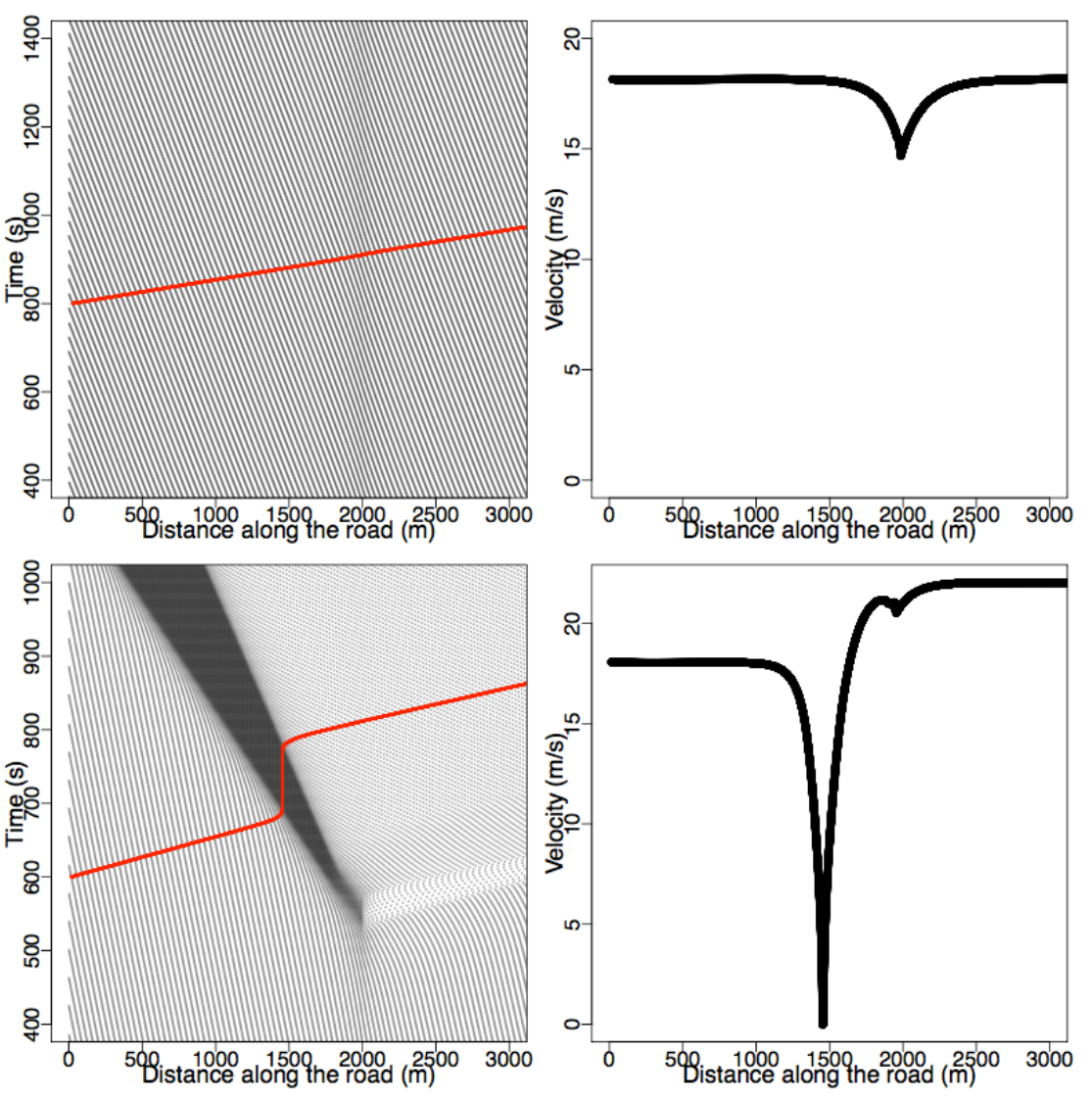}
\caption{The spatio-temporal representation of the traffic flow similar to the ones in Fig.(\ref{phase}). Here the red line is the trajectory of one single vehicle as it passes through the intersection. The plot on the right shows the velocity of this tracked vehicle (on the left) as a function of its position along the road. The top row shows the motion of a vehicle passing through the intersection in the free flow phase, while the bottom row shows the motion of the vehicle in the congested phase.}
\label{velocity}
\end{center}
\end{figure}

In the free flow phase vehicles decelerate when they approach the intersection, and there is a dip in the vehicle velocity only when they are close to the intersection. In general the decrease of the velocity is also relatively small. In the congested phase, however, wide moving jams form and move upstream. The velocities of the vehicles within the wide moving jam drop almost to zero (see Fig.(\ref{velocity})). The characteristic out-flow $F_{dj}$ downstream of a wide moving jam\cite{kernerbook} is smaller than $F_{\text{in}}$, leading to the growing of the width of the wide moving jam. 

\begin{figure}[h!]
\begin{center}
\includegraphics[height=7cm]{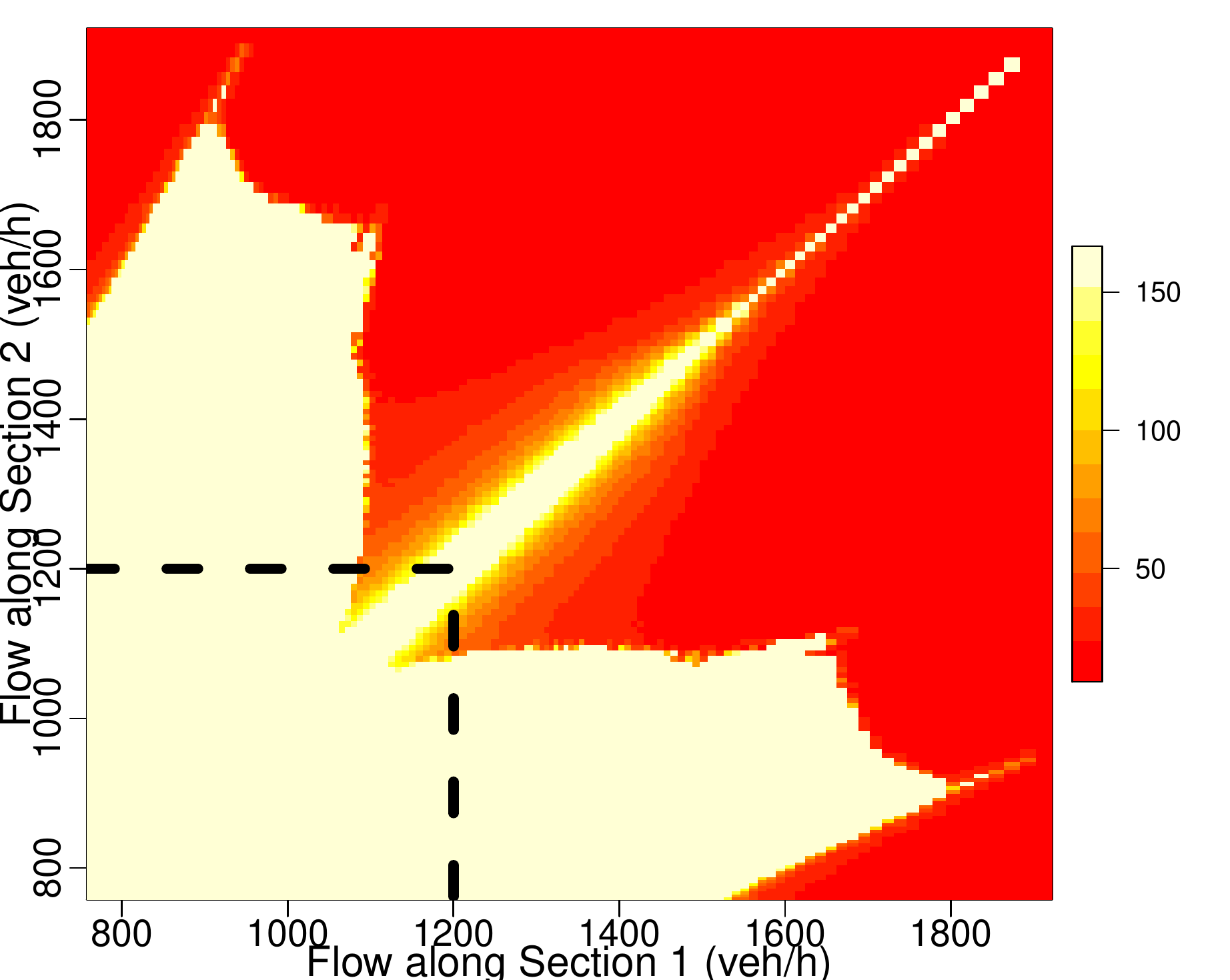}
\caption{The color scale here shows the time it takes for the congestion to develop because of the intersection. Note the white region shows no congestion after numerical simulation of more than three hours, indicating the free flow phase. The black dotted lines indicate the theoretical upper limit of the intersection control capacity with the traffic light control, which we explain in more details in Sec.~\ref{tlimit}.}
\label{twophase}
\end{center}
\end{figure}

When the density of the traffic (or the in-flow) is within the range of the absolutely stable phase, wide moving jams cannot occur, since $F_{\text{in}}<F_{dj}$ will lead to the shrinking of the jam width. The condition $F_{\text{in}}>F_{dj}$, however, does not always lead to the congested phase, since in the metastable phase it depends on whether the effective perturbation induced by the intersection is large enough\cite{bando,kerner}. Numerical simulation shows the phase boundary depends on the in-flow of both the road section in a non-trivial way (see Fig.(\ref{twophase})).

The simulation of Fig.(\ref{twophase}) is performed by assuming constant average in-flow of both the two road sections, with the vehicles being inserted randomly. Though it is not shown in the plot, one should note that the traffic is always in the free flow when the $F_{\text{in}}$ of both the road intersection is smaller than $\sim 1200\text{veh/h}$. As will be discussed in Sec.~\ref{op_com}, in this particular case the capacity of the intersection with any signalized control has an upper bound of $\sim 1200\text{veh/h}$. Fig.(\ref{twophase}) illustrates the superior performance of the intersection capacity. One should also note the interesting fact that when the inflows of the two road sections are similar, the congested phase do not occur even for very high inflows. This ``lock-in" effect deserves further attentions and will be discussed elsewhere.
\begin{figure}[h!]
\begin{center}
\includegraphics[height=4cm]{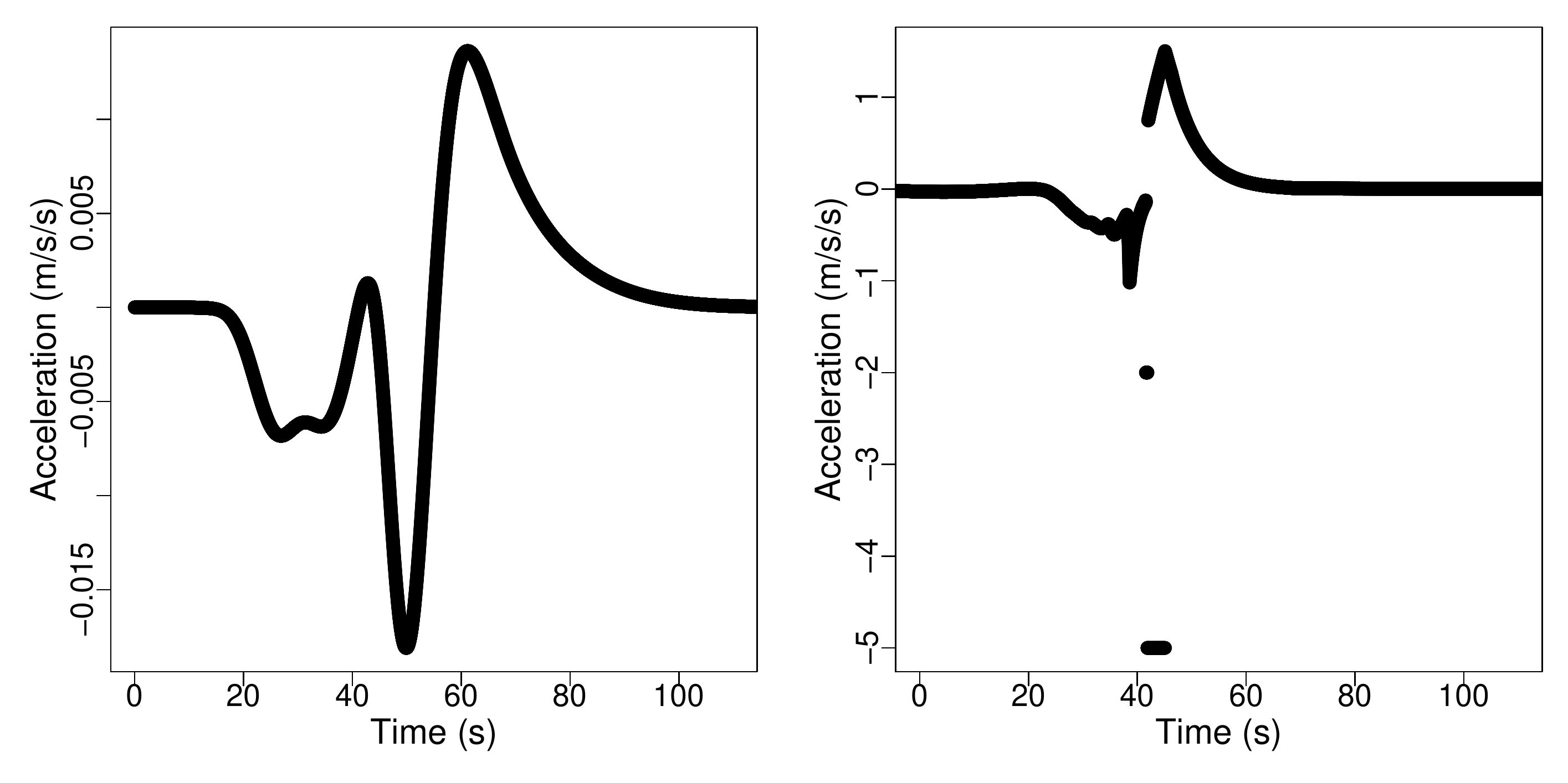}
\caption{The two typical acceleration profiles of individual vehicles close to the intersection. On the left the vehicle is almost unhindered with small acceleration/decelerations when approaching the intersection. This is very common when the inflow is relatively small, highlighting the efficiency of the control algorithm. On the right, emergency brake in the caution is triggered for less than 5 seconds, and this could occur when the inflow is close to the intersection capacity.}
\label{figacc}
\end{center}
\end{figure}

For individual vehicles passing through the intersection, our numerical simulation shows their acceleration as a function of time is rather smooth, with no rapid and large oscillations (see Fig.(\ref{figacc})). One would expect, however, that such oscillations can occur in more complicated situations or in real-world implementations. Smoothening of such oscillations without compromising the intersection control efficiency as well as as the issue of safety could be necessary, and we discuss this issue in more details in Sec.~\ref{op_com}.
\begin{figure}[h!]
\begin{center}
\includegraphics[height=6.5cm]{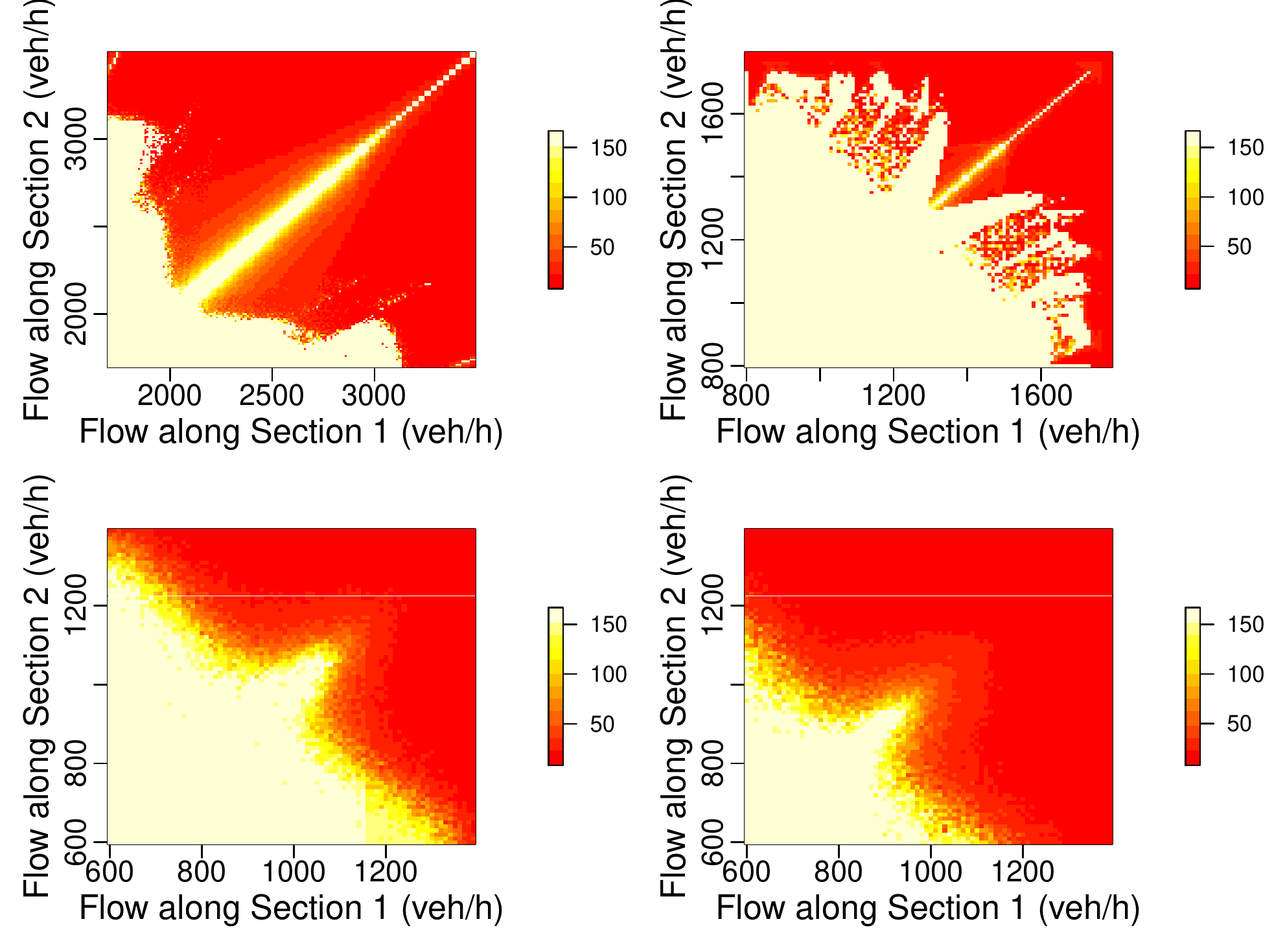}
\caption{We repeated the same numerical simulation as Fig.(\ref{twophase}). Top left: we replace the traffic model tuned in this paper by the IDM\cite{idm}. The parameters of the IDM are not fitted with the empirical data of the traffic flow in Singapore, but with the highway traffic flow in Germany. Top right: The phase diagram of the intersection with two lanes in each direction. Bottom left: the phase diagram of the intersection with a single lane in each direction, and with turning allowed. Here around 10\% of the vehicles turn at the intersection in each lane. Bottom right: Same numerical simulation as the one in Bottom left, but with 20\% of vehicles turning at the intersection in each lane.}
\label{figturn}
\end{center}
\end{figure}

The basic intersection control algorithm is applicable for more complicated intersection configurations with multi-lanes or with turnings allowed. It also works with various different traffic models proposed in the literature (i.e. the intelligent driver model\cite{idm} or the three phase traffic models\cite{threephase}, see Fig.(\ref{figturn})). While the overall interaction-based control scheme is universal, one should also note that with more complex intersection, additional fine tuning is required to ensure efficiency and especially the robustness. In particular, the parameters in Table. II can be adjusted, and Eq.(\ref{deceleration}) can also be generalized. We include some examples in Fig.(\ref{figturn}). When turning is allowed, we also need to specify the turning velocity which should be much lower than the speed limit of the road. Here we set the maximum velocity when the turning vehicle approaches the intersection to be $10ms^{-1}$. Further research is in progress for more detailed studies of the different types of intersection configurations.

\subsection{Mixed Traffic Flow}

We will also show numerical evidences that our intersection control is robust when there is a mixture of vehicles with and without ICC devices. We set the policy that if a vehicle does not have an installed ICC device, it has to come to a full stop at the intersection before moving on (the intersection effectively is governed by a STOP sign, since the traffic light is absent). This is obviously a simple but very crude way to accommodate the vehicles without the ICC device. We use it to show that even without much effort in optimising the policies for the mixed traffic, our scheme can still accommodate it with reasonably good performances.

\begin{figure}[h!]
\begin{center}
\includegraphics[height=8.5cm]{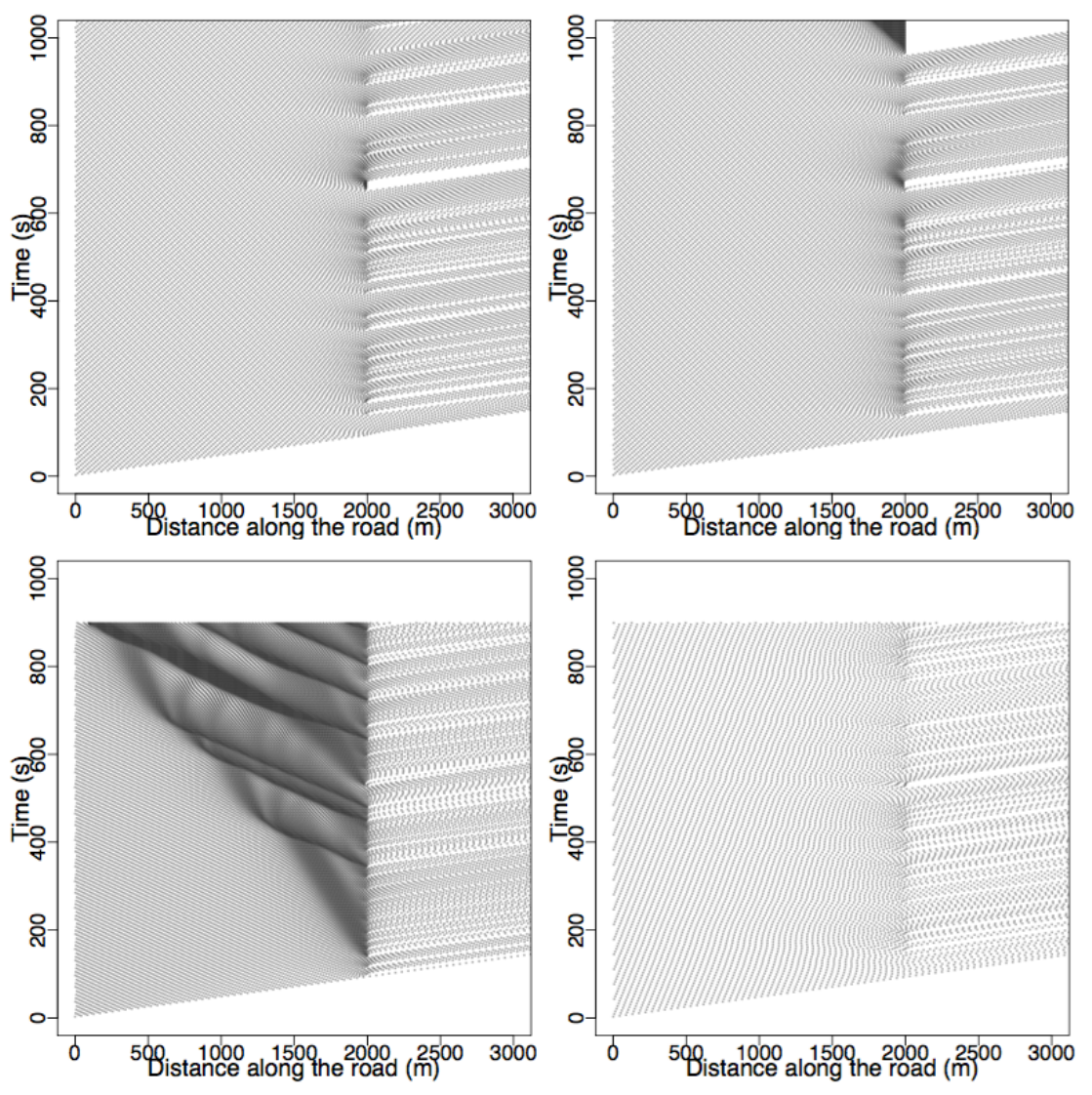}
\caption{The spatio-temporal representation of the mixed traffic flow. The top two plots show the mixed flow in the free flow phase for both the road sections, when the in-flow of the traffic is $1100 \text{veh/h}$. The bottom two plots show the mixed flow for both the road sections. Road section 1 has an in-flow of $1400 \text{veh/h}$, and congestion develops (the left plot). Road section 2 has an in-flow of $750 \text{veh/h}$, and the free flow is maintained. In all cases, a random distribution of $10\%$ of vehicles do not have the ICC device installed.}
\label{mixed}
\end{center}
\end{figure}

The first step is to model this human behaviour properly. The car following model of Eq.(\ref{human}) is sufficient to model such human drivers in most of the cases. In addition to Eq.(\ref{human}), in our numerical simulation the ``STOP" sign at the intersection requires such vehicles to uniformly decelerate once they come close to the intersection, until they come to a complete stop right in front of the intersection. The vehicle will start to cross the intersection only when the vehicle approaching the intersection in the other road section is sufficiently far away. Quantitatively, this happens if it takes at least three seconds for the vehicle in the other road section to reach the intersection at its current velocity. This is considered safe, because not only is ``three seconds" sufficient enough for the vehicle to cross the intersection from zero velocity, the vehicle in the other road section (either human drivers or vehicles equipped with ICC device) is also expected to slow down in all situations.

One should note that vehicles with the ICC device installed follow the algorithm Eq.(\ref{i1}$\sim$\ref{i4}) to avoid collisions at the intersection, based on the information relayed by the beacon at the intersection. Such information only includes the positions and velocities of the relevant vehicles as described in Fig.(\ref{schematics}), and it does not matter if the relevant vehicles are equipped with the ICC devices or not. Thus our algorithm will make sure the ICC equipped vehicles can make decisions about passing through the intersections or giving way to the waiting vehicles without the ICC device at the intersection in an optimal manner. 

\begin{figure}[h!]
\begin{center}
\includegraphics[height=6.5cm]{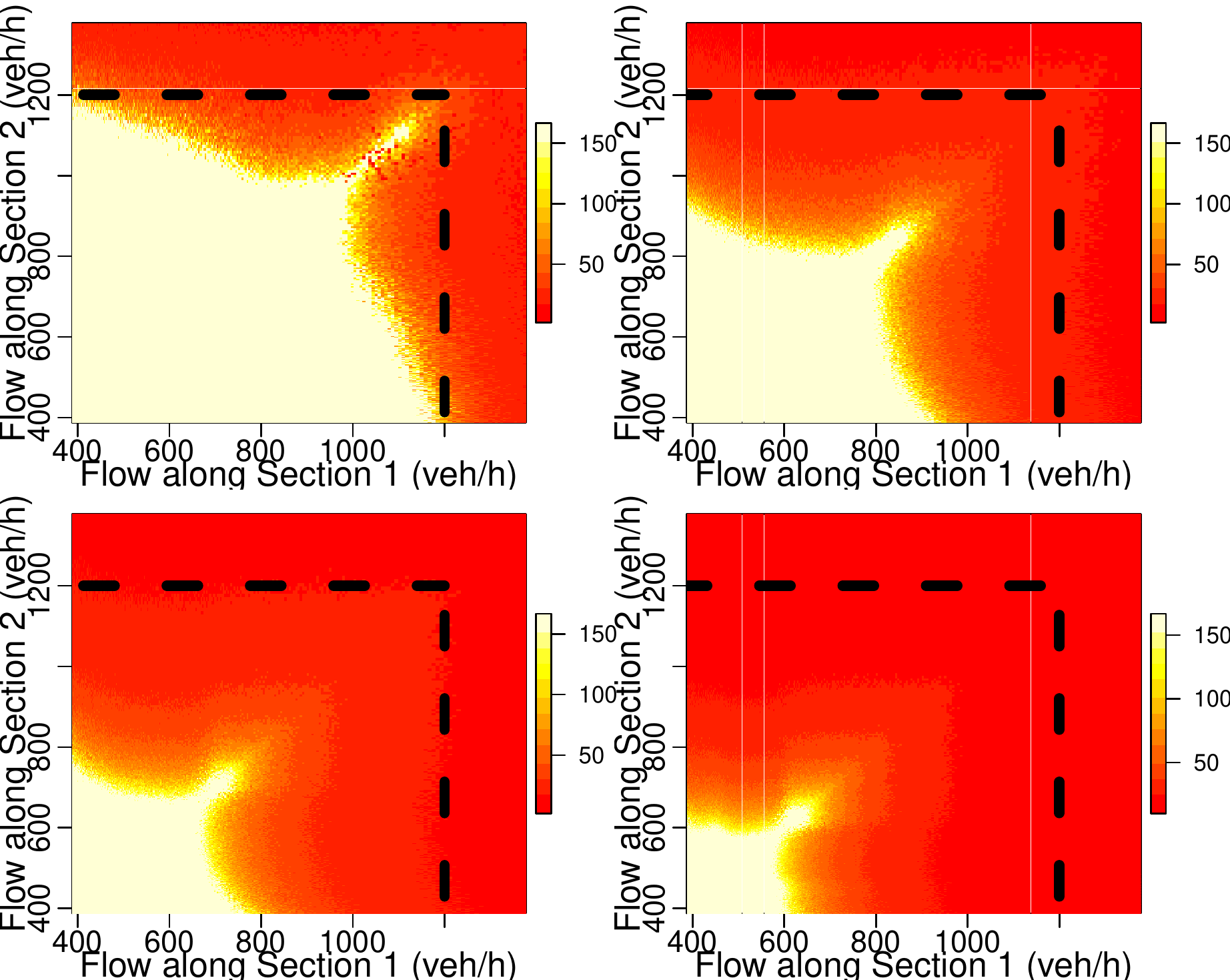}
\caption{Phases of the traffic when there is a mix of vehicles with and without the ICC device installed. The color scale is the time it takes for the congestions to evolve, and the white color indicates no congestion after more than three hours of traffic with the constant inflow. The percentages of human drivers (vehicles without ICC device) are $10\%$ (Top left), $30\%$ (Top right), $50\%$ (Bottom left) and $70\%$ (Bottom right). The black dotted lines are the same as the ones in Fig.(\ref{twophase}), which we explain in more details in Sec.~\ref{tlimit}.}
\label{twophasemixed}
\end{center}
\end{figure}

Requiring vehicles to make a full stop right before crossing the intersection significantly decreases the intersection capacity. This can be illustrated by the numerical simulation assuming constant in-flow in both the road sections, but now with a certain percentage of vehicles without ICC devices. The intersection capacity is now significantly reduced, as one can see in Fig.(\ref{twophasemixed}).

The numerical simulation shows that the intersection control efficiency is on par with the traffic light control even if the ICC penetration rate is only $\sim 30\%$. One should note, however, the simplicity of our traffic modelling and simulation, especially with vehicles not equipped with the ICC device, implies the quantitative analysis of the control algorithm should only indicate a general trend of the efficiency of the control. The simulated intersection capacity also does not reflect the fact that the trip across the intersection is rather inefficient for vehicles without the ICC device. The simulations in Fig.(\ref{twophasemixed}) are intended to show strong evidence that our lightless intersection control algorithm can accommodate mixed traffic flow, with the expected decrease of the control efficiency when the proportion of vehicles without the ICC device increases.

It is thus important to increase the proportion of vehicles installed with the ICC device, especially at intersections with busy traffic. On the other hand, preliminary numerical simulation shows the time it takes for the congestions to evolve depends weakly on the number of lanes: in many cases, as long as the inflow \emph{per lane} is low, one can double the number of lanes, and thus the total inflow of the road section, without significantly increasing the chance of the occurrence of the congested phase (see Fig.(\ref{figturn})). More detailed studies of the multi-lane road system are still in progress and will be presented elsewhere.

\section{Safety, Driver Comfort and Efficiency}\label{op_com}

The numerical analysis from the previous section shows strong evidence that the proposed scheme is robust and collision free, even though a very simple model for the control algorithm is used. For actual implementations, many factors need to be considered. First and foremost, the safety of the vehicles around the intersection has to be guaranteed, even for rare situations such as disruptions, accidents and rogue drivers. Secondly, the overall driving experience, now punctuated by the ICC device, should still be smooth and comfortable; lastly, the efficiency of the lightless intersection control should be superior as compared to the signalised control.

While we expect the accommodation of all these factors can be achieved by generalising the simple model for the control algorithm (e.g. Eq. (\ref{deceleration})) under the general concept we proposed, a full optimisation with hardware specifications and actual implementation is beyond the scope of this paper. In this section, on the other hand, we will illustrate the capability of our design in guaranteeing safety and balancing between comfort and efficiency, with the systematic tuning of a few important parameters in the algorithm.

\subsection{Tuning Parameters in the Algorithm}

In Table. II, the parameters $V_m, d_s$ and $d_c$ are in principle tunable numerically; however in reality the values of these parameters are also constrained by the road configuration, comfort level of the drivers, as well as other various physical limits. Thus the tuning of these parameters should take into consideration of a wide variety of factors, instead of just focusing on the optimization of the intersection control alone.

The parameter of $L_s$, on the other hand, is intrinsic to our algorithm. For a single intersection it can be tuned freely and affecting only the performance of the intersection control. From the robustness point of view, we would like $L_s$ to be as large as possible so there is ample space for the vehicles to synchronize their intersection time headway. The larger the value of $L_s$, the less the need for the vehicles to hard brake in the caution zone. This is desirable both from the improvement of the comfort level and from the energy conservation point of view.

On the other hand, from the practical point of view we would also like $L_s$ to be small, because drivers within the synchronization zone have to follow the traffic rules of the intersection control, rather than driving at will. The operational range of the beacon is also limited, and the distance between two consecutive intersections may also be short. These are also the factors to be considered when determining the optimal value of $L_s$.

The safety parameters $l_{\text{safe}}$ and $t_{\text{safe}}$ are the two most important tuning parameters in the algorithm, in addition to the generalisation of Eq. (\ref{deceleration}). Tuning of these two parameters offer an intuitive and systematic way in enhancing the safety and comfort level of the traffic flow, while at the same time still maintaining the efficiency of the intersection control. 

It is clear that both $l_{\text{safe}}$ and $t_{\text{safe}}$ should be kept as small as possible, if efficiency is concerned. They are both implicitly the measure of the strength of perturbation induced by the intersection: if both of them are set to be zero, the effects of the intersection virtually disappear, and the two road sections become decoupled. The values of $l_{\text{safe}}$ and $t_{\text{safe}}$ are, however, constrained by the width of the intersection, lengths of the vehicles as well as the response time and precisions of the devices. The larger the safety parameters, the greater the gaps between the vehicles at the intersection, therefor enhancing the safety margin of the intersection control, making it more robust against accidents and malfunctions. The values of these two parameters are also strongly related to the comfort level of the passengers in the vehicles, as we will show in details below.

\subsection{Individual Vehicle Dynamics}

In Sec.~\ref{numerics_phase} the vehicles are inserted at the beginning of the traffic lane at random positions with fixed average in-flow. Such stochasticity tends to attenuate when the vehicles move towards the intersection, as in our simulations the traffic model is deterministic with identical drivers. To fully investigate the issue of safety and robustness of the intersection control algorithm in the presence of stochasticity and unruly driving behaviours, we first study the dynamics of a small number of vehicles passing through the intersection, with various different initial conditions in the interaction zone. These initial conditions include both the common situations occurring frequently in the empirical traffic flow, and rare situations when drivers make minor infractions such as exceeding the speed limit, approaching the intersection with insufficient reduction of speed, or accidental stops near the intersection. 

Since the dynamics of the vehicles are governed by the non-linear interactions between vehicles crossing the intersection, it is not possible to predict analytically how the vehicles will decelerate when they approach the intersection. We look at the case of two single vehicles, as well as the case of two small platoons of vehicles, approaching the intersection from different directions. 
\begin{figure}[h!]
\begin{center}
\includegraphics[height=9cm]{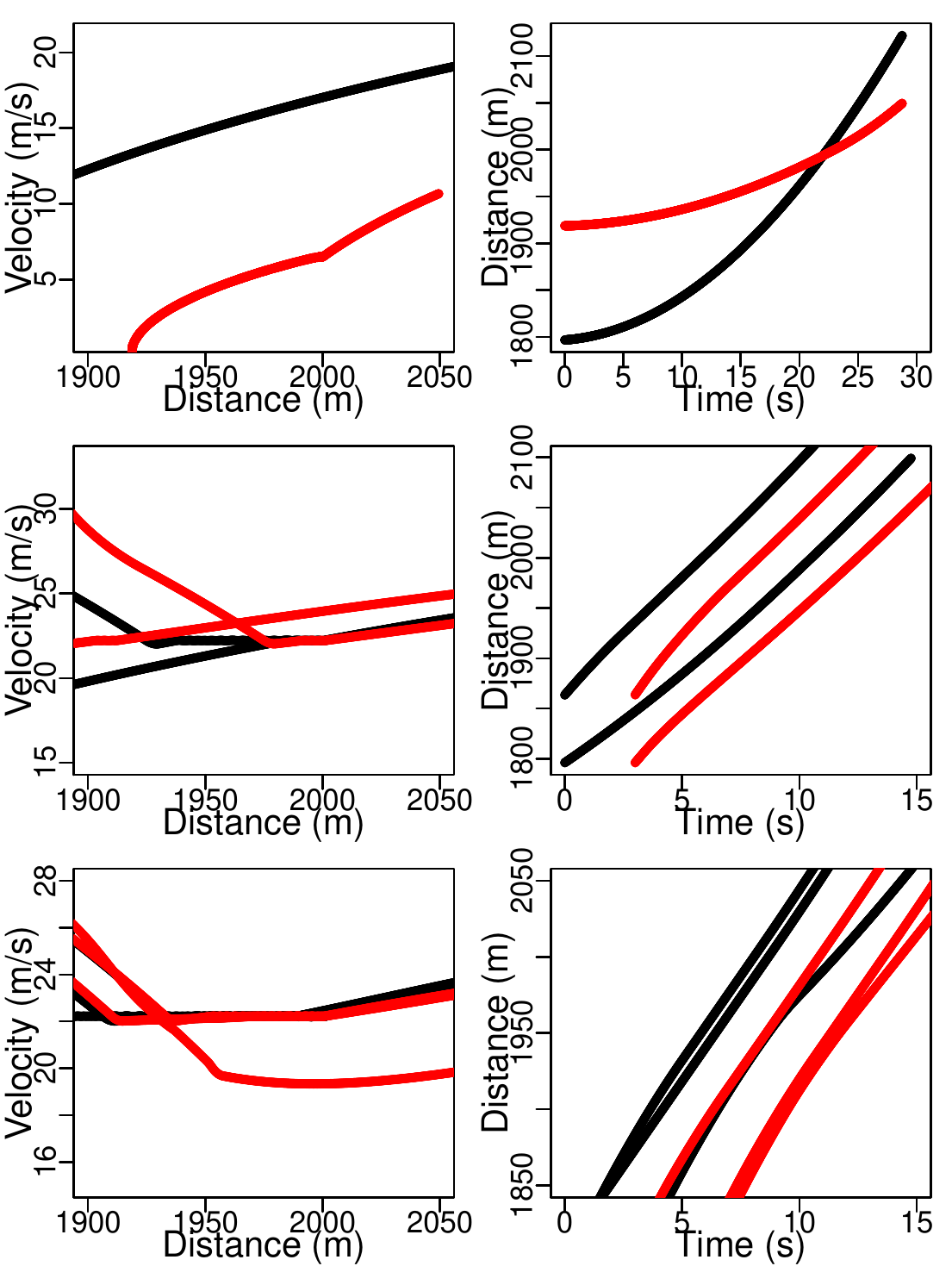}
\caption{Trajectories of individual vehicles approaching the intersection. The distances are measured along the direction of travel for both lanes meeting at the intersection, which is located at $2000 m$. The synchronization zone starts at $203.7 m$ away from the intersection, as determined by the ICC induced deceleration of $2ms^{-2}$ in the zone. The caution zone starts at $50 m$ away from the intersection, as determined by the ICC induced deceleration of $5ms^{-2}$ in the zone. In both cases, the zone is long enough so that the vehicles can come to a complete stop from the speed limit with the assigned deceleration. The black plots are for vehicles travelling from west to east, while the red plots are for vehicles travelling from south to north. The top row is for a single vehicle in each direction approaching the intersection, with the vehicle travelling from south to north starting at zero velocity in the synchronization zone. The second row is for a platoon of two vehicles in each direction approaching the intersection, while the third row is for a platoon of three vehicles in each direction approaching the intersection; in both these two cases, the vehicles enter the synchronization with different but relatively high velocities. }
\label{trajectory}
\end{center}
\end{figure}

As can be clearly seen from Fig.(\ref{trajectory}), if the vehicles enter the intersection with high velocity, they will generally decelerate in a self-organised way, avoiding each other at the intersection. If the vehicles enter the intersection with very small velocity, they will accelerate, but vehicles from one direction tends to accelerate less in order for the vehicles from the other directions to pass first. For the platoon of vehicles, the interactions at the intersection involving the leading vehicles are sufficient in guiding the entire fleet through the intersection.

\begin{figure}[h!]
\begin{center}
\includegraphics[height=6cm]{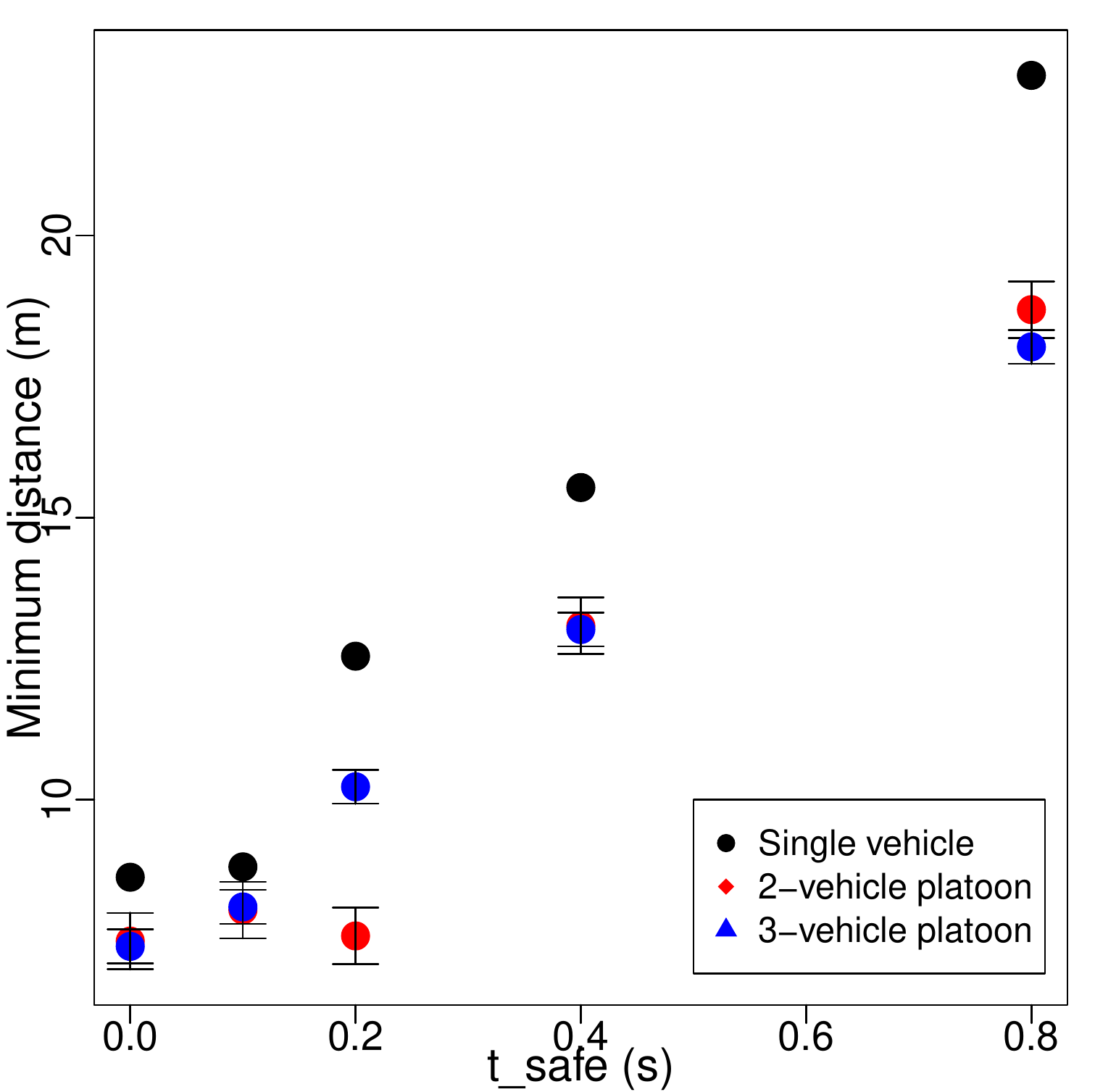}
\caption{The minimum distances between the vehicle just about to enter the intersection, from the vehicles passing the intersection, obtained from different initial conditions with $l_{\text{safe}}=0, t_{\text{safe}}$ given by the x-axis. The different initial conditions are scanned from velocities ranging from zero to $33 ms^{-1}$ and positions ranging from $203.7 m$ (the start of the synchronization zone) to $50m$ (the start of the caution zone) away from the intersection. For the platoon of vehicles, the time gaps between consecutive vehicles travelling in the same direction has a random fluctuation, since vehicles are inserted stochastically to mimic the empirical observations. This leads to the error bars in the plot.}
\label{gap}
\end{center}
\end{figure}

For a continuous traffic flow across the intersection , we can verify numerically that with larger $l_{\text{safe}}$ and $t_{\text{safe}}$, the gap between vehicles at the intersection is also greater, implying more secure passage across the intersection (see Fig.(\ref{gap})). 

\begin{figure}[h!]
\begin{center}
\includegraphics[height=6cm]{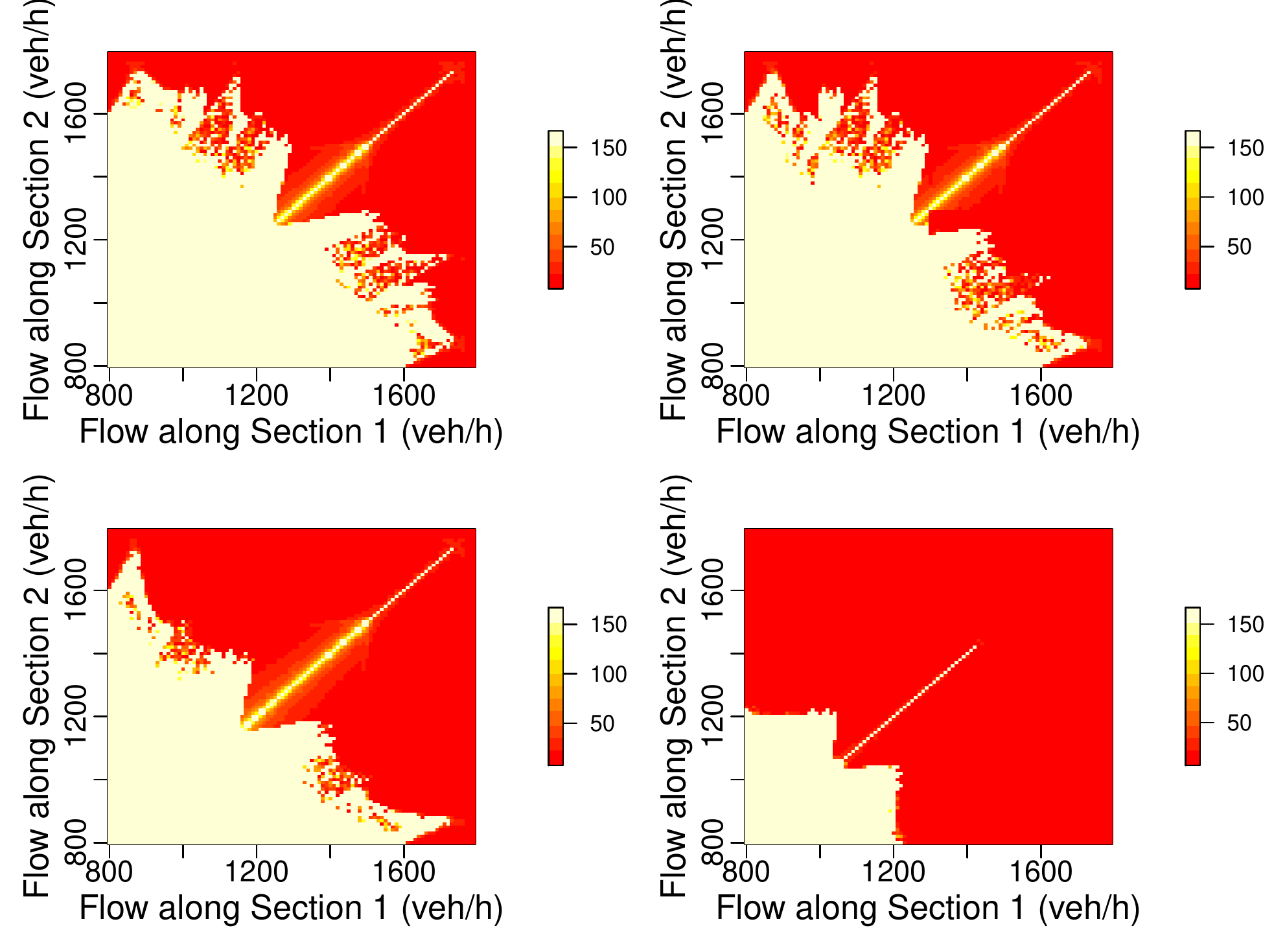}
\caption{The four plots are the same numerical simulation of the lightless intersection control with different values for the safety parameters. Top left: $l_{\text{safe}}=9 m,t_{\text{safe}}=0.2s$; Top right: $l_{\text{safe}}=11 m,t_{\text{safe}}=0.2s$; Bottom left: $l_{\text{safe}}=9 m,t_{\text{safe}}=0.4s$; Bottom right: $l_{\text{safe}}=9 m,t_{\text{safe}}=0.8s$. The numerical simulations are performed by random insertion of the vehicles while keeping the average inflow fixed.}
\label{safetyphase}
\end{center}
\end{figure}

From an optimization point of view, we thus would like to minimize both $l_{\text{safe}}$ and $t_{\text{safe}}$, while at the same time guaranteeing the safe passage of vehicles across the intersection. We evaluate the intersection capacity of the control by inserting in vehicles continuously into the traffic system, and monitoring the maximum inflow, beyond which the traffic congestions will occur (see Fig.(\ref{safetyphase})). 

The compromise between the measure of safety and the measure of the intersection control efficiency is clearly seen from a few case studies in Fig.(\ref{gap}) and Fig.(\ref{safetyphase}), when $l_{\text{safe}}$ and $t_{\text{safe}}$ are tuned. One should note that increasing $l_{\text{safe}}$ and increasing $t_{\text{safe}}$ are not equivalent, given the control algorithm as defined by Eq.(\ref{i1}) $\sim$ Eq.(\ref{i3}). While both increase the time gap on the RHS of the inequalities, the effect of $l_{\text{safe}}$ also depends on the vehicle's velocity. 

\subsection{Driver Comfort}

\begin{figure}[h!]
\begin{center}
\includegraphics[height=6cm]{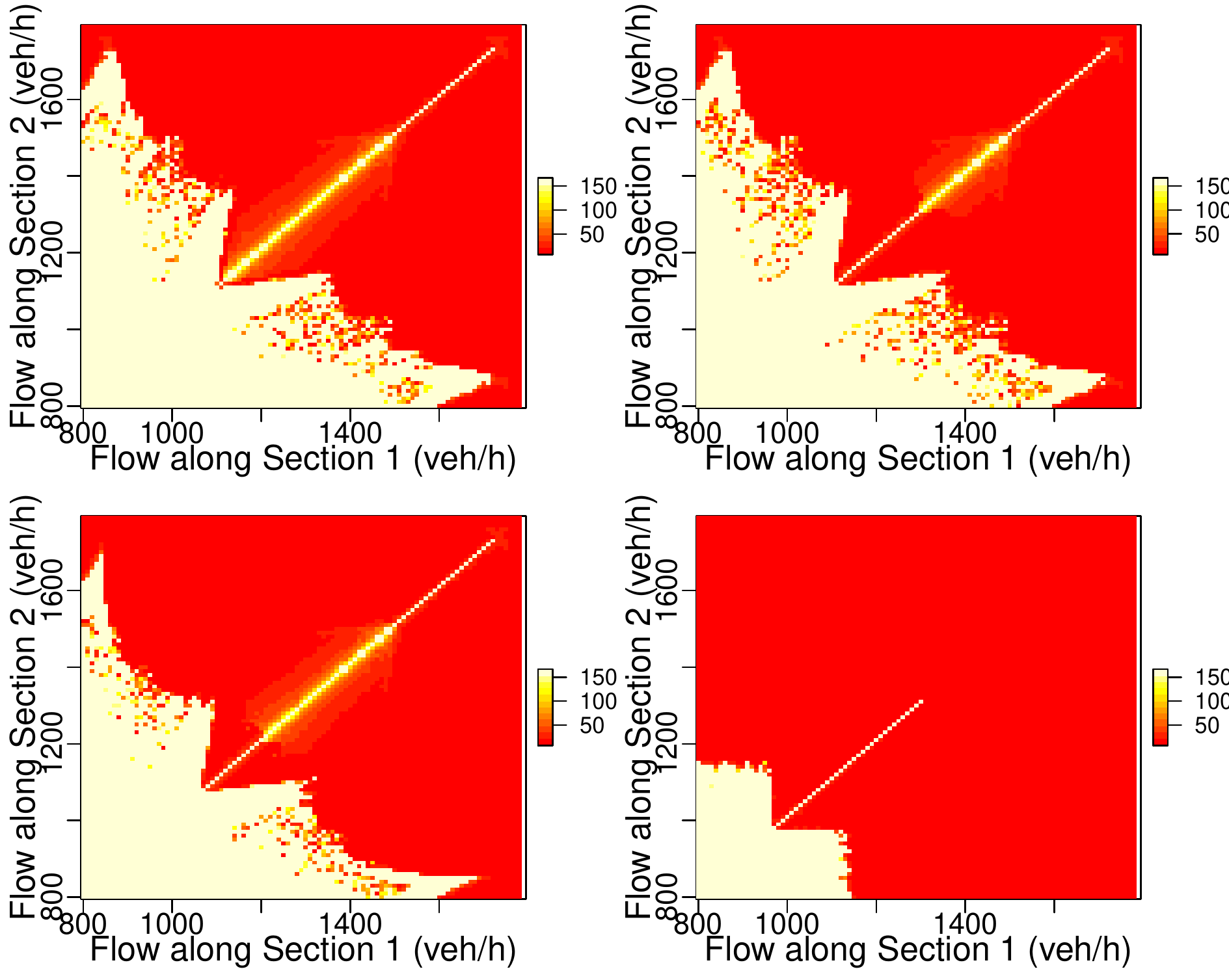}
\caption{The same plot as the one in Fig.(\ref{safetyphase}), but with the jerk satisfying Eq. (5), and with different safety parameters. Top left: $l_{\text{safe}} = 9m$, $t_{\text{safe}} = 0.3s$; Top right: $l_{\text{safe}} = 11m$, $t_{\text{safe}}=0.2s$; Bottom left: $l_{\text{safe}}=9m$, $t_{\text{safe}}=0.4s$; Bottom right: $l_{\text{safe}}=9m$, $t_{\text{safe}}=0.8s$. The numerical simulations are performed by random insertion of the vehicles while keeping the average inflow fixed.}
\label{jerkphase}
\end{center}
\end{figure}

All previous simulations in the paper do not consider the comfort level of the drivers. In particular, when the ICC decelerates the vehicle, the acceleration changes abruptly with very large jerk. We can remedy this by enforcing a comfortable acceleration profile as discussed in\cite{jerk}. A large jerk (as long as it is not too large) is not necessarily uncomfortable, as long as the acceleration changes in a rather smooth and predictable way. We implement the sinusoidal acceleration profile studied in\cite{jerk}, and the resulting phase diagram is shown in Fig.(\ref{jerkphase}).

Formally, when the ICC device activates the dynamical control, it actually institutes an acceleration profile of: 
\begin{eqnarray}\label{jerkprofile}
a\left(t\right)=a\left(0\right)+\frac{\Delta a}{2}\left(1\pm\cos\left(\frac{\pi}{\Delta t}t\right)\right)
\end{eqnarray} 
Where $a\left(0\right)$ and $a\left(\Delta t\right)$ are the accelerations before and after the big change instituted by the ICC device. This could happen at the moment when the ICC device start to decelerate the vehicle, or the moment the ICC device stops deceleration and the human driver is back in control by following the vehicle in the front. $\Delta a = a\left(\Delta t\right)-a\left(0\right)$. The time interval $\Delta t$ is chosen so that the maximum absolute value of the jerk will not exceed $20ms^{-3}$. Such restrictions imply vehicles do not reach the desired acceleration instantaneously, affecting the capability of avoiding a potential collision. This, however, can be compensated by increasing $l_{\text{safe}}$ and $t_{\text{safe}}$, at the cost of reducing the efficiency of the intersection control.
     The implementation of the jerk profile of Eq.(\ref{jerkprofile}), while increasing the driver comfort level, does reduce the efficiency of the intersection control, as the capacity of the intersection control is reduced. For $l_{\text{safe}}=9m$ and $t_{\text{safe}}=0.2s$, implementing Eq.(\ref{jerkprofile}) will cause a $0.01\%$ chance of collisions at the intersection, while instantaneous change of the acceleration is completely safe as far as numerical simulation is concerned. For $l_{\text{safe}}=9m$ and $t_{\text{safe}}=0.4s$, $l_{\text{safe}}=9m$ and $t_{\text{safe}}=0.8s, l_{\text{safe}}=11m$ and $t_{\text{safe}}=0.2s$, no collisions occur even if the jerk is constrained by Eq.(\ref{jerkprofile}). However, the red region (congested phase) is slightly larger in Fig.(\ref{jerkphase}), as compared the corresponding plots in Fig.(\ref{safetyphase}).

\subsection{Theoretical Limits of the Signalized Control}\label{tlimit}

We also would like to know how well this algorithm is compared to the traffic light control for a single intersection. With the traffic light to control the intersection traffic, the red light period will invariably bring one section of the traffic to a complete stop, when the vehicles are queuing and waiting for the green light. The maximum capacity of the intersection is thus independent of the details of the schemes as long as the traffic light is involved. The flow downstream of the queuing vehicles when the traffic light turns from red to green is the same as the flow downstream of a wide moving jam, and is characteristic of the driving behaviours of the traffic dynamics\cite{kernerbook}. Thus the capacity of the intersection with the signalized control is bounded by this characteristic flow; in our particular example in Sec.~\ref{numerics} this characteristic flow is $\sim 1200 veh/h$, which is obtained from empirical observations and fitted by the numerical model of Eq.(\ref{ovm}).

One should also note that in practice, the actual intersection capacity of the signalized control is much lower than the theoretical limit of $1200 veh/h$ in our example (see the black dotted lines in Fig.(\ref{twophase}) and Fig.(\ref{twophasemixed})). This is because with the red light, the outflow of the traffic in that direction is zero, further reducing the average outflow of the intersection. When the inflow of the two road sections are similar, by symmetry every section will have an equal chance of the red light no matter what signalized control scheme is used. Thus for this particular case the intersection capacity is $\sim 600 veh/h$, worse than our scheme even if $70\%$ of the traffic is made of conventional human drivers. Similary, even with relatively large $l_{\text{safe}}$ and $t_{\text{safe}}$, and relatively high comfort level, the intersection capacity with lightless intersection control is always much superior as compared to the signalized traffic control (see Fig.(\ref{jerkphase})). 

When the intersection capacity is not reached, each vehicle passing through the intersection only suffer a short period of velocity depression. The minimum velocity occurs close to the intersection with the value around $80\%$ of the prevalent velocity when the vehicle is far away from the intersection (see Fig.(\ref{velocity})). Combined with the fact that the intersection capacity is generally much smaller for signalized intersection control, it is thus obvious the effective travel time for most cases should be much shorter with the scheme proposed in this paper.

\section{Limitations and Practical implementation}\label{op_limit}

Our algorithms have the potential to be scaled and generalised for more sophisticated intersections, and we will include a detailed analysis elsewhere, with data from the empirical tests. We have shown, however, that the simple interaction based algorithm works well for several different intersection types, including multilane traffic and turning lanes. It also works well with conventional vehicles mixed with vehicles equipped with the ICC device.

\begin{figure}[h!]
\begin{center}
\includegraphics[height=8cm]{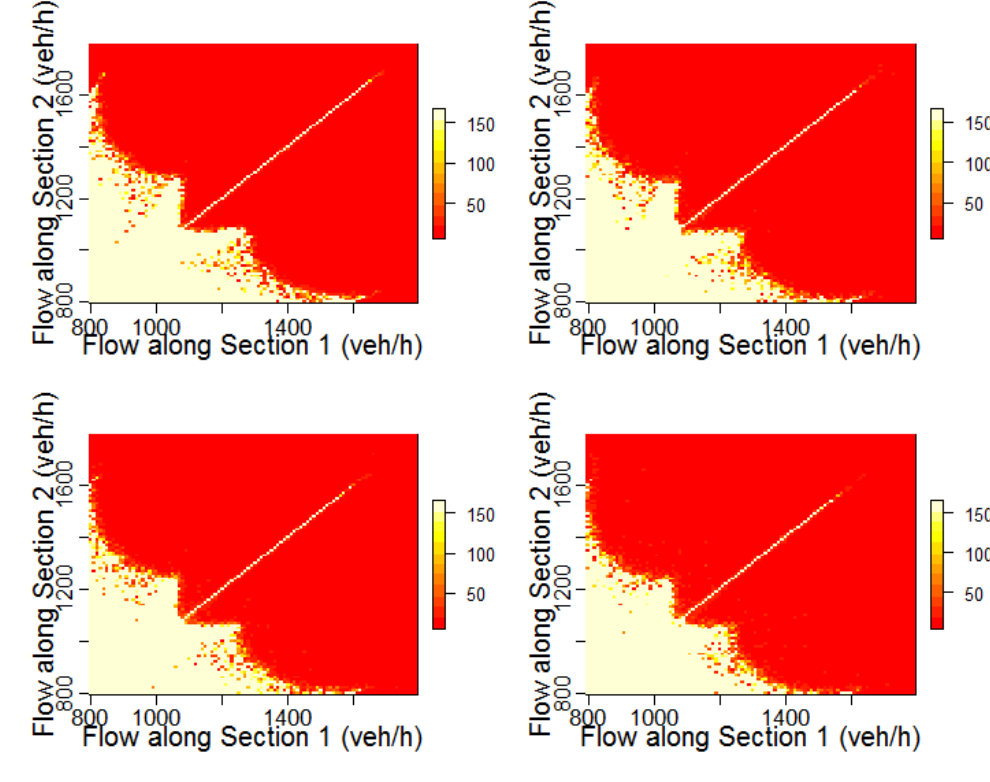}
\caption{The same plot as the one in Fig.(\ref{jerkphase}) with the jerk satisfying Eq. (5) and with two types of vehicles. The small vehicles with length $5m$ have $l_{\text{safe}}=9m$, and the large vehicles with length $10m$ have $l_{\text{safe}}=14m$. The other safety parameter, $t_{\text{safe}}=0.3s$. The two types of vehicles are mixed randomly. Top left: $10\%$ of large vehicles; Top right: $15\%$ of large vehicles; Bottom left: $20\%$ of large vehicles; Bottom right: $25\%$ of large vehicles.}
\label{figphase_t}
\end{center}
\end{figure}

From the perspective of modelling and numerical simulations, one should note that any such endeavour is an approximation of the reality. While the simulations show strong evidence of the viability of the concept we are proposing, the quantitative aspects especially with regard to the intersection control efficiency should only be understood as a reference to the performance of our lightless intersection control algorithm. In particular, we have assumed identical vehicles in our numerical simulations, while in reality vehicles of different sizes and driving behaviours are present on the road in a stochastic manner. Our algorithm can be easily adapted for such scenarios. The simplest option is to increase both $l_{\text{safe}}$ and $t_{\text{safe}}$; one can choose, for example, $l_{\text{safe}}$ to be greater than the sum of road width and the maximum vehicle length that can occur on the road. This will make sure gaps between vehicles are large enough for all types vehicles, at the cost of significantly reducing the intersection control efficiency. A better approach is to make $l_{\text{safe}}$ vehicle specific, so that for each ICC device, the safety parameters are determined by the physical characteristics of the vehicle on which it is installed. In Fig.(\ref{figphase_t}), we mix vehicles of length $5m$ and vehicles of length $10m$ randomly on the road, with different $l_{\text{safe}}$ for the two vehicle species. 

The presence of large vehicles results in greater effective perturbation of the traffic flows at the intersection, explicitly in the form of the increased $l_{\text{safe}}$ for the large vehicles. It is interesting also to see that the phase diagram is only weakly dependent on the percentage of large vehicles on the road, at least when such vehicles are still a minority. The random distribution of the large vehicles also smoothen out regions of ``no-congestions" in the phase diagram, which could be an artefact of the assumption of identical vehicles. The simulations in Fig.(\ref{figphase_t}) are more realistic, and one can further improve it by adding more vehicle species and by employing more realistic and stochastic traffic models. A more extensive study on these issues will be presented elsewhere.

We would also like to add a few comments here about a rather interesting feature appearing in most of the intersection capacity phase diagrams, in which the free flow persists even for very high inflow (e.g. $\sim 1800 veh/h$ per section in Fig.(\ref{twophase})), as long as the inflows along both road sections are very close to each other, leading to a ``sharp needle" along the diagonal of the phase diagram. For example, a significant decrease of inflow along one road section from $1800 veh/h$, while keeping the other road section inflow at $1800 veh/h$, will paradoxically reduce intersection capacity and lead to congestions. This rather unrealistic simulation result is due to the simplicity of the deterministic traffic models with identical drivers, and is probably related to the instabilities described in Ref.\cite{gonzales}. The introduction of stochasticity in Fig.(\ref{figphase_t}) attenuates such ``sharp needles"; with the mix of human drivers in Fig.(\ref{twophasemixed}), they can disappear entirely. In both cases more than one species of vehicles appear in the traffic system, which is the more realistic scenario. We would also expect such artefact to disappear when the inflows of the traffic fluctuates in time, as one would expect for the real traffic systems.

Our scheme currently cannot accommodate pedestrians or cyclists crossing the roads, so it may not be applicable to all road intersections. In places where there are few pedestrians or cyclists, the lightless intersection control can be accompanied by the auxiliary traffic light for non-motorised travellers, at the cost of the occasional suspension of the lightless intersection control and the lowering of the efficiency. For places where pedestrians and cyclists are common, it may be worthwhile for overpasses or underpasses to be constructed, if the benefits of the lightless intersection control are large enough (in particular if the traffic volume is large).

For conventional vehicles approaching the intersection, we require them to come to a full stop at the intersection. This effective ``stop sign" is not optimal for the conventional vehicles, and it also lowers the overall efficiency of the intersection control. The lightless intersection control is thus more appropriate if the majority of the vehicles on the road are equipped with the ICC device. The inconvenience for the vehicles without the ICC device could also been seen as an incentive for car owners to upgrade their vehicles. Here, we just need to illustrate that our algorithm will not break down if there are conventional vehicles, or vehicles with broken ICC devices.

With the lightless intersection control one invariably needs to have new traffic rules; however we can try to mirror the conventional traffic rules for them to be easily understood and accepted. First of all, it is compulsory for vehicles equipped with ICC to have the device turned on once the vehicle is within the interaction zone. It could be configured such that the ICC is turned on automatically when the vehicle enters the interaction zone. This is analogous to the conventional scenario in which it is compulsory for drivers approaching the intersection to obey the signals as long as the traffic light is on.

By design the driver is in full control of the vehicle throughout the trip. While the active ICC will control the acceleration or deceleration of the vehicle, it can be overridden anytime by the driver. However, such voluntary action is analogous to the conventional scenario in which the driver decides not to obey the traffic light. Without valid reasons this will be treated as a violation of the traffic law.
\begin{figure}[h!]
\begin{center}
\includegraphics[height=2.5cm]{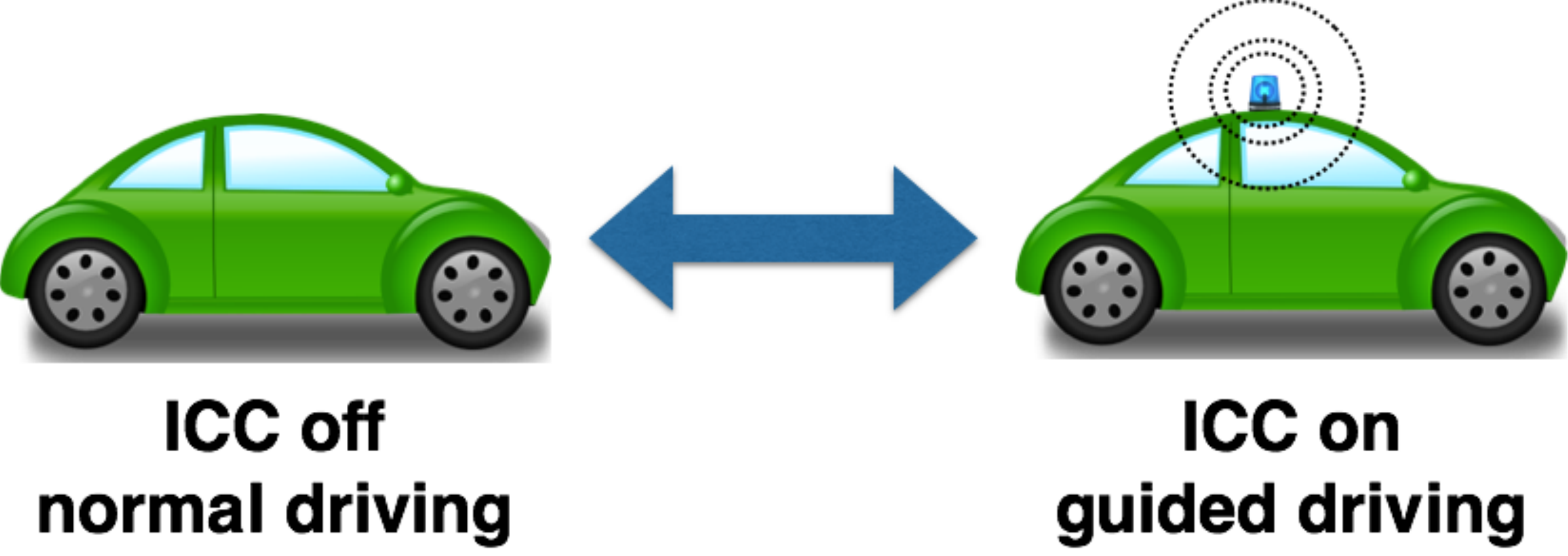}
\caption{Whether the ICC device is active should be visually discernible. When the vehicle enters the interaction zone the ICC device should be automatically switched on. After passing the intersection the device will be automantically switched off. The flashing of the light bulb on the top of the car on the right is just a schematic representation to show that whether or not the ICC is active can be visually distinguished.}
\label{icc}
\end{center}
\end{figure}

It is thus important that whether the ICC is turned on can be visually distinguished by the external parties. This can be easily achieved by a certain light signal pattern on the vehicle (for example, the intermittent flashing of the brake light).  Fellow drivers as well as the traffic police can thus easily inspect if the traffic laws are obeyed. In case of dispute or accidents, the fact of which party override the ICC improperly should be the key in determining who is at fault.  

Additional traffic rules may be needed for more complex intersections, and we will study this in more details elsewhere. As an example, while our algorithm can be easily generalised to multi-lane traffic (see Fig.(\ref{figturn})), we do require vehicles in the interaction zone not to change lanes, since in the interaction zone the ICC devices are activated. The interaction zone is a rather short stretch of the road from the intersection, and lane-changing should be rare and considered as bad driving behaviours by convention. While allowing lane-changing is possible, it requires more complex algorithms and hardwares for the lightless intersection control with little to gain. 

\section{Summary and Outlook}\label{summary}

In summary, we have proposed a comprehensive scheme of the intersection control with no traffic lights, by assuming that the majority of the vehicles on the road are equipped with an ICC device that can guide the vehicles through the intersection. The scheme includes algorithms for such intersection control, which takes into account of the vehicle-to-vehicle interaction both within the road section and in between different road sections. Unlike most of the lightless intersection control scheme introduced in the literature, in our proposal the architecture and algorithm of the intersection control is much simpler and completely decentralized. This is important not only from the cost, scalability and security point of view, but also for the inclusion of conventional vehicles driven entirely by humans. Conceptually, it is also good to know what the minimal level of ``intelligence" is required of vehicles for highly efficient lightless intersection control to be possible. We have shown that a disruptive improvement of the intersection control does not require all vehicles to be autonomous (driverless), and the functional requirement of the ICC device is rather basic and achievable with mature technologies. Such intersection control can also coexist with intersections with traffic lights, and be implemented for all or selected intersections in the urban area. 

While the ICC device by default controls the acceleration or braking of the vehicle when it is on, it does not, however, control the steering of the vehicle; the driver also can stop the ICC device at any time. The fact that the driver is always in control of the vehicle can potentially make the adaptation of the lightless intersection control much easier, since it closely resembles the conventional driving experience. The ICC device, together with the beacon at the intersection, plays the analogous role of the traffic light: while its control over the vehicle is passive, it carries the legal authority. Thus policy-wise it is also convenient to transition from the conventional signalized intersection control to the new scheme we proposed.

Our numerical simulation, which takes into account of the full interactions between vehicles, shows the scheme is robust and safe, accommodating many of the practical constraints mentioned above. With our capability to realistically simulate very high traffic flow at the intersection, we can demonstrate that our scheme is highly efficient as compared to the conventional signalized intersection control. Though the performance of the intersection control is significantly affected by mixing vehicles without ICC devices, our scheme is fully capable for the mixed traffics, which is essential for it to be implemented in the near future.

The architecture of our scheme allow easy upgrades to more advanced algorithms for the lightless intersection control, such as the ones already proposed in the literature. In our scheme from the hardware point of view, the collection and distribution of the information is centralized, while the control of the vehicle dynamics is fully distributed. Additional layers of more centralized control from the beacon, especially for certain rare occasions, can be achieved without compromising too much the virtue of the distributed dynamical control. From the algorithmic point of view, only a very small amount of vehicles are involved in the interaction around the intersection. While we have shown the efficacy of this level of involvement, with better hardware capabilities one can involve more vehicles further away from the intersection. We also have shown that surprisingly simple interaction rules are sufficient for very busy intersections, but more complicated rules can be considered if they significantly improve the performance of the intersection control.

The new paradigm introduced by our scheme is a natural candidate that not only cater to the immediate need for traffic efficiency in crowded cities, but also evolve smoothly into the age when autonomous vehicles play a dominant role in our daily lives. The work presented here is just the first step in justifying the feasibility of the scheme we introduced, with the aim of achieving disruptive improvement in terms of traffic efficiency, while at the same time being compatible with both the conventional and emerging next generation traffic infrastructure, with low technological barrier and ease of adaptation. Work is ongoing for more detailed studies of multi-lane, two-way traffic that allows turning, as well as improvement in both the efficiency and the robustness of the algorithm. Preliminary investigation suggests that our algorithm is highly scalable, as the intersection capacity mainly depends on the inflow per lane, instead of the total inflow. Extension into the lightless intersection control of the urban road network is also underway; the instability towards traffic congestions due to the effective perturbations of the traffic flow from the network of intersection can also be an interesting topic of research\cite{gonzales}. Prototype experimentations as well as empirical data are needed to quantify the time and energy efficiency of our scheme, especially in comparison with other lightless intersection control schemes involving driverless vehicles in the literature.

\section{Acknowledgement}
This research was partially supported by Singapore A*STAR SERC ``Complex Systems" Research Programme grant 1224504056. We would like to thank Prof. Lynette Cheah from Singapore University of Technology and Design and Prof. Jitamitra Desai from Nanyang Technological University for useful comments.


\begin{thebibliography}{99}
\bibitem{footnote}
By ``semi-automated" we mean vehicles with driver assistance systems such as adaptive cruise control\cite{lunge,montanaro,jerath}, or other types of devices that advise or control the dynamics of the vehicle during special circumstances.
\end{thebibliography}
\end{document}